\documentclass[preprint,12pt,nonatbib]{elsarticle}

\usepackage{amssymb}
\usepackage{amsmath}
\usepackage{upgreek}
\usepackage{algorithmicx}
\usepackage{algorithm}
\usepackage{algpseudocode}
\usepackage{hyperref}
\usepackage{graphicx}
\usepackage{booktabs} 

\DeclareMathOperator*{\argmin}{arg\,min}
\makeatletter
\let\c@author\relax
\makeatother
\usepackage[style=numeric, backend=biber,sorting=none]{biblatex}
\journal{Journal of Computational Physics}

\begin{document}

\begin{frontmatter}

\title{Feature-Guided Sampling Strategy for Adaptive Model Order Reduction of Convection-Dominated Problems}

\author[UC]{Ali Mohaghegh } 
\ead{ali.mohaghegh@ku.edu}

\author[UC]{Cheng Huang} 
\ead{chenghuang@ku.edu}

\affiliation[UC]{organization={University of Kansas},
            city={Lawrence}, 
            state={KS},
            country={USA}}

\begin{abstract}
Though high-performance computing enables high-fidelity simulations of complex engineering systems, accurately resolving multi-scale physics for real-world problems remains computationally prohibitive, particularly in many-query applications such as optimization and uncertainty quantification. Projection-based model order reduction (MOR) has demonstrated significant potential for reducing computational costs by orders of magnitude through the creation of reduced-order models (ROMs). However, physical problems featuring strong convection, such as hypersonic flows and detonations, pose significant challenges to conventional MOR techniques due to the slow decay of Kolmogorov N-width present in these problems. In the past few years various approaches have been proposed to address this challenge; one of the promising methods is the adaptive MOR. In this work, we introduce a feature-guided adaptive projection-based MOR framework tailored for convection-dominated problems involving flames and shocks. This approach dynamically updates the ROM subspace and incorporates a feature-guided sampling method that strategically selects sampling points to capture prominent convective features, ensuring accurate predictions of crucial dynamics in the target problems. We evaluate the proposed methodology using a suite of challenging convection-dominated test problems, including shocks, flames, and detonations. The results demonstrate the feature-guided adaptive ROM's capability in producing efficient and reliable predictions of the nonlinear convection-dominated physical phenomena in the selected test suite, which are well recognized to be challenging for conventional ROM methods.
\end{abstract}

\begin{keyword}
adaptive model reduction, proper orthogonal decomposition, hyper-reduction, convection-dominated problems, feature-guided sampling
\end{keyword}

\end{frontmatter}

\section{Introduction}

While rapid advances in computing technologies are making high-fidelity simulations feasible for complex engineering systems \cite{wang_towards_2017,aditya_direct_2019,oefelein_advances_2019}, directly embedding these simulations in many-query workflows such as optimization and uncertainty quantification remain impractical due to their high computational costs. In this regard, projection-based model order reduction (MOR) \cite{lumley_low-dimensional_1997,benner_survey_2015} has emerged as a promising technique to construct reduced-order model (ROM) from the full-order model (FOM) for these high-fidelity simulations via rigorous mathematical reduction. As demonstrated by many researchers \cite{carlberg_galerkin_2017,mcquarrie_data-driven_2021,huang_model_2022,groth_computational_2006}, the resulting ROM inherits the majority of the FOM's modeling capabilities while achieving orders of magnitude acceleration in computational time. Typically, the construction of projection-based ROMs involves two main stages: (1) an \emph{offline} stage that performs a limited number of FOM simulations over a range of target parameters and compute low-dimensional subspaces to closely approximate the high-dimensional FOM solutions; (2) an \emph{online} stage that constructs and executes ROM by projecting the FOM equations onto the computed low-dimensional subspace. Classical projection-based MOR often leverages linear subspaces such as those provided by proper orthogonal decomposition (POD) for ROM construction. However, linear subspaces present ineffectiveness in representing convection-dominated problems (common characteristics in most multi-physics and multi-scale problems such as combustion, hypersonics and detonations) featuring slow decay of the Kolmogorov N-width. which is often referred to as the Kolmogorov barrier \cite{bonnaillie-noel_efficient_2016,cohen_optimal_2020}. During the past few years, several remedies have been proposed to break this barrier and construct more effective ROM for convection-dominated problems, which can be categorized into four main groups:

\begin{enumerate}
    \item The first group of researchers resorts to using \emph{multiple} local linear subspaces to represent the high-dimensional dynamics. Each local subspace is tailored for a specific region of the state space in the target problems. Different from the conventional approaches using a \emph{single} global subspace for the entire state space, these local subspaces offer significant improvements in dimensionality reduction and lead to substantial savings in ROM construction, which would otherwise require an impractical number of basis modes via a single global subspace. The local subspaces are often computed by clustering the state space into distinct groups, each of which corresponds to a unique feature in the target problems. These features can be defined based on parameter space \cite{eftang_hp_2010,eftang_parameter_2012,peherstorfer_localized_2014,geelen_localized_2022}, time \cite{rapun_reduced_2010,parish_windowed_2021,shimizu_windowed_2021}, predefined similarity metrics such as k-means \cite{peherstorfer_localized_2014,amsallem_fast_2015,amsallem_nonlinear_2012,grimberg_mesh_2021,kaiser_cluster-based_2014}, or dissimilarity criteria such as projection error \cite{amsallem_pebl-rom_2016}.

    \item The second group of researchers aims to overcome the Kolmogorov barrier by leveraging advanced nonlinear manifolds to obtain more accurate low-rank approximation of convection-dominated physics. Some of researchers in this group adopt explicit quadratic nonlinearity in approximating the FOM solutions \cite{rutzmoser_generalization_2017,barnett_quadratic_2022,geelen_operator_2023} or compute nonlinear manifolds directly using convolutional autoencoders \cite{lee_model_2020,kim_fast_2022}. These approaches offer a promising alternative for capturing complex dynamics that are inadequately represented by linear subspaces. More recently, efforts have been made to use an artificial neural network (ANN) to describe the low-dimensional subspaces for construction of projection-based ROM \cite{barnett_neural-network-augmented_2023, halder_reduced-order_2024}, which can effectively reduce the dimensionality of the online approximation of the solution beyond the limits achievable with affine and quadratic approximation manifolds, while preserving accuracy.

    \item The third group of researchers pursue transformations of either the subspaces or the snapshots to improve the capabilities of linear subspaces in representing convenction-dominated dynamics by aligning parameters or time-dependent structures through appropriate transformations. A prominent class of approaches leverages transformations of subspaces or snapshots to construct these manifolds, with the goal of recovering low-rank structures exhibiting fast Kolmogorov N-width decays, such as freezing methods \cite{ohlberger_nonlinear_2013}, transported subspaces or snapshots \cite{nair_transported_2019}, and implicit feature tracking \cite{alireza_mirhoseini_model_2023}. Among all these methods, one popular approach is to use the shifted POD (sPOD) based on the concept of transport compensation, which aims at enhancing the approximation of a linear description by aligning the parameters or time-dependent structures with the help of suited transforms as more data collected \cite{fedele_symmetry_2015,karatzas_projection-based_2020,krah_front_2023,vermolen_model_2021,rowley_reduction_2003,welper_transformed_2020,taddei_space-time_2021,rim_manifold_2023,krah_robust_2024}. In addition, symmetry reduction has been used leveraging symmetries, e.g., translation invariance of the underlying partial differential equations (PDEs), in combination with a POD-based reduction to reduce dimensionality of problem by exploiting their inherent symmetries \cite{fedele_symmetry_2015,rowley_reduction_2003}. This method was shown to be a specific instance of sPOD in \cite{black_projection-based_2020}. Another approach employs transported subspaces, explicitly using hyperbolic PDE characteristics or tracking the reduced system's front \cite{krah_front_2023,krah_robust_2024,rim_manifold_2023}. A third approach uses space-time registration to align the subspace with local flow features \cite{taddei_space-time_2021}. In the work by Welper \cite{welper_transformed_2020}, transformed snapshots replace standard POD and produce comparable results if proper alignment is performed. Nonino et al. \cite{nonino_reduced_2023} apply invertible parameter-dependent mappings of the flow domain to enable parameter-dependent deformations in ROM.

    \item The fourth group of researchers leverage adaptive MOR by updating the linear basis on the fly during the online ROM calculations to achieve an optimal representation of the \emph{local} target dynamics \cite{brand_fast_2006,sapsis_dynamically_2009,peherstorfer_online_2015,singh_lookahead_2023,zucatti_adaptive_2023,iollo_advection_2014,huang_predictive_2023,rumpler_adaptive_2014,peherstorfer_model_2020}. One approach is to use a hybrid snapshot simulation with an on-the-fly criterion to determine when to use FOM instead of a local ROM when POD modes are inadequate or unrepresentative for making accurate predictions. Bai et al. \cite{bai_deim_2020,bai_reduced_2022} devised criteria based on the sufficiency of the reduced basis to determine when to switch between FOM and a local ROM, with rapid low-rank singular value decomposition (SVD) updates. Feng et al. \cite{feng_fomrom_2021} formulated a rigorous \emph{a posteriori} error estimator to control switching between FOM and ROM. Yano et al. \cite{yano_globally_2021} developed a trust-region method that informs the ROM construction process to meet accuracy requirements in topology optimization problems. Carlberg \cite{carlberg_adaptive_2015} introduced a method for enriching the reduced-basis space online by splitting a basis vector into multiple vectors with disjoint support, similar to mesh-adaptive h-refinement. This approach was later improved by Etter and Carlberg \cite{etter_online_2020} through vector-space sieving. Another approach focuses on exploring the FOM to estimate crucial information, which is then used to update the low-rank linear subspaces. Peherstorfer \cite{peherstorfer_model_2020} adapted the affine approximation space by exploiting the spatiotemporal locality of propagating coherent structures derived from sampled FOM solutions. This method was recently extended to convection-dominated physics problems in premixed flame simulations \cite{uy_reduced_2022}. Some of recent works from Babaee’s group \cite{patil_reduced-order_2023,jung_accelerating_2025,ramezanian_--fly_2021} directly derived evolution equations for low-rank structures using variational principles to construct an on-the-fly ROM method for species transport in chemically reacting flow applications.

\end{enumerate}

Though the aforementioned four groups of methods show potentials to break the Kolmogorov barrier, one remaining challenge lies in the resulting ROMs' restricted predictive capabilities in describing the convection-dominated physics that are different from the ones used to train the ROM during the offline stage and there is usually no guarantee regarding the accuracy of the ROM predictions \cite{camacho_investigations_2024}. In this regard, the adaptive MOR discussed above opens a promising avenue to address this limitation by minimizing \cite{peherstorfer_model_2020}, or completely eliminating \cite{ramezanian_--fly_2021}, the offline training stage requirement while resorting to efficient online exploitation of FOM to build ROM on the fly, which inherently enhances the predictive capabilities of the ROM and enables true predictions of convection-dominated features in the problems \cite{huang_predictive_2023}. Though promising, one crucial area for improvement in adaptive MOR formulation lies in the selection and adaptation of sparse samples (1) to evaluate and approximate the nonlinear terms in nonlinear ROM construction (often referred to as hyper-reduction or sparse sampling), and (2) to exploit the FOM equation residuals to update the low-dimensional subspaces. One common approach to achieve this in adaptive MOR \cite{naderi_adaptive_2023,peherstorfer_breaking_2022} is to directly leverage conventional sparse sampling approaches such as gappy POD \cite{everson_karhunenloeve_1995} and discrete empirical interpolation method (DEIM) \cite{chaturantabut_nonlinear_2010}. In addition, more systematic sparse sampling methods \cite{chaturantabut_nonlinear_2010,yuxiang_beckett_zhou_model_2012,carlberg_gnat_2013,zimmermann_geometric_2018,peherstorfer_stability_2020} have been developed and explored in the literature, which can be potentially applied in the adatpvie MOR formulation. However, we remark that most of these methods seek to minimize the interpolation error associated with sparse sampling for improved stability and robustness of the resulting ROM, which does not necessarily guarantee accurate representation of important FOM physics, especially those featuring strong convection of sharp gradients or discontinuities. 

In the present work, leveraging the previous work by Huang and Duraisamy \cite{huang_predictive_2023} as well as Camacho and Huang \cite{camacho_investigations_2024}, we develop a feature-guided adaptive projection-based MOR formulation for convection-dominated problems featuring sharp gradients (e.g., flames) and discontinuities (e.g., shocks and detonations). Specifically, we devise a feature-guided sampling method that strategically guide the selection of sparse samples based on the most prominent convective features in the target problems, to achieve optimal efficiency and effectiveness in the resulting adaptive ROMs. We perform detailed evaluations of the proposed method based on a suite of challenging convection-dominated test problems, which includes (1) Sod shock tube, (2) freely propagating laminar flame and (3) multi-species detonation tube.

The remainder of this paper is organized as follows. Section \ref{sec:fom} and \ref{sec: std_MOR} present the formulation of the FOM and conventional ROM respectively. Section \ref{sec:AROM} introduces the feature-guided adaptive MOR formulation. Section \ref{sec:algorithm} discusses the computational algorithms and complexity of the proposed adaptive MOR formulation. Section \ref{sec:results} presents numerical results of test cases with assessment of their robustness and efficiency. Section \ref{sec:conclusion} provides concluding remarks and perspectives.

\section{Full-Order Model}
\label{sec:fom}
We first represent the full-order model (FOM) as a generic dynamical system

\begin{equation}
\frac{\text{d}\mathbf{q}}{\text{d}t} = \mathbf{f}(\mathbf{q}, t) \quad \text{with} \quad \mathbf{q}(t = 0) = \mathbf{q}_0 , 
\label{eq:fom}
\end{equation}

\noindent where \( t \in [t_0, t_f] \) is the solution time, and \( t_0, t_f \in \mathbb{R}_{\geq 0} \) are the initial and final solution time instances, respectively. In the context of time-dependent PDEs, the above dynamical system represents the spatially-discretized system. For PDEs describing conservation laws, \( \mathbf{q}  \in \mathbb{R}^N \) is referred to as the conservative state, where \( N \) is the total number of degrees of freedom in the system. $\mathbf{q}_0 \in \mathbb{R}^N $ is the vector of states to be specified as the initial conditions at t = 0. Similarly, the non-linear function \( \mathbf{f}: \mathbb{R}^N\to \mathbb{R}^N \) represents the spatial discretization of any flux, source, and body force terms in the conservation system, as well as any boundary conditions. The number of degrees of freedom may be decomposed as \( N = N_{var} \times N_{elem} \), where \( N_{elem} \) is the number of points in the spatial discretization of the original PDE form of the governing equations (e.g., the Navier-Stokes equations), and \( N_{var} \) is the number of state variables being solved for at each spatial point (e.g., density, momentum, total energy). Various time-discretization methods can be introduced to solve Eq. \ref{eq:fom}, such as implicit or explicit multi-step or Runge-Kutta methods. In the current work, we use the explicit four-stage third degree strong stability preserving Runge-Kutta (four-stage SSP-RK3) method \cite{durran_numerical_2010}. Any RK method in a standard form can be expressed as following:

\begin{equation} 
\mathbf{q}^{n} = \mathbf{q}^{n-1} + \Delta t \sum_{i=1}^{s} \beta_i \textbf{f}(\mathbf{q}^{(i)}, t^{(i)}),
\label{eq:SSPRK_grouped}
\end{equation}

\noindent where intermediate states $\mathbf{q}^{(i)}$ can be found using

\begin{equation}
\label{eq:SSPRK_intem}
\mathbf{q}^{(i)} = \mathbf{q}^{n-1} + \Delta t \sum_{j=1}^{i-1} \alpha_{ij} \textbf{f}(\mathbf{q}^{(j)}, t^{(j)}),
\end{equation}

\noindent the coefficients $\alpha_{ij}$ and $\beta_{i}$ are determined based on applied time-integration method \cite{durran_numerical_2010} , and $\Delta t$ is the the physical time step for numerical solution.

\section{Standard Static Model Order Reduction}
\label{sec: std_MOR}

In this section, we briefly review the formulation for standard model order reduction (MOR).

\subsection{Low-dimensional subspace for solution variables}
\label{subsec:std_MOR}

Any static reduced order model is developed using a dataset composed of snapshots collected at various time instances, referred to as $\mathbf{q'} \in \mathbb{R}^{N}$. The data matrix, with each of its columns representing a solution snapshot $\mathbf{q'}$, is denoted as $\mathbf{Q}$. Next, a low-rank approximation of the FOM state, $\mathbf{q}$, in Eq. \ref{eq:fom} is computed, 

\begin{equation}
\mathbf{\tilde{q}}(t) = \mathbf{q_{\text{ref}} + P^{-1}V{q}}_r(t),
\label{std. main rom}
\end{equation}

\noindent where $\mathbf{{\tilde{q}}}\in \mathbb{R}^{N}$ is the approximation of $\mathbf{q}$. The reference state,$\mathbf{q_{\text{ref}}}$, is commonly applied in fluid dynamics problems, and possible reference states include the initial solution, $\mathbf{q_{\text{ref}}} = \mathbf{q}(t = t_0)$, or the time-averaged solution, $\mathbf{q_{\text{ref}}} = \frac{1}{\Delta T} \int_{t_0}^{t_0 + \Delta T} \mathbf{q}(t) \, dt$. $\mathbf{V} \in \mathbb{R}^{N \times n_p}$ is the trial basis matrix, $\mathbf{P} \in \mathbb{R}^{N\times N}$ is a scaling matrix and $\mathbf{q}_r\in \mathbb{R}^{n_p}$ represents the reduced state with $n_p$ representing the number of trial basis modes. In the current work, $\mathbf{V}$ is computed via the proper orthogonal decomposition (POD) \cite{berkooz_proper_1993,lumley_structure_1967} from the singular value decomposition (SVD), which is a solution to

\begin{equation}
\label{eq:svd}
\argmin_{V \in \mathbb{R}^{N \times n_p}} \| \mathbf{PQ} - \mathbf{VV}^T \mathbf{PQ} \|_F \quad \text{s.t.} \quad \mathbf{V}^T \mathbf{V} = I.
\end{equation}

\noindent The scaling matrix, $\mathbf{P}$, is applied to $\mathbf{q'}$ so that the variables corresponding to different physical quantities in the data matrix $\mathbf{Q}$ have similar orders of magnitude. Otherwise, $\mathbf{Q}$ may be biased by physical quantities of higher magnitudes (e.g., total energy). In this work, we normalize all quantities by their $L^2$ norm, as proposed by \cite{lumley_low-dimensional_1997}:

\begin{equation}
\mathbf{P} = \text{diag}(\mathbf{P}_1, \dots, \mathbf{P}_i, \dots, \mathbf{P}_{N_{\text{elem}}}),
\end{equation}

\noindent where $\mathbf{P}_i = \text{diag}(\phi_{1, \text{norm}}^{-1}, \dots, \phi_{N_{\text{var}}, \text{norm}}^{-1})$. Here, $\phi_{v, \text{norm}}$ represents the $v$-th state variable and is defined as

\begin{equation}
\phi_{v, \text{norm}} = \frac{1}{\Delta T} \int_{t_0}^{t_0 + \Delta T} \frac{1}{\Omega} \int_{\Omega} \phi'_v(x, t)^2 \, dx \, dt.
\end{equation}

\subsection{Galerkin projection}

Model reduction using the Galerkin projection is formulated for the continuous-time representation of the FOM as given in Eq. \ref{eq:fom}. This involves scaling Eq. \ref{eq:fom} with the scaling matrix $\textbf{P}$, followed by projecting it onto the test space \(\mathbf{V}\)

\begin{equation}
\label{prj.rom}
\mathbf{V}^T \mathbf{P} \frac{\text{d} \mathbf{q}}{\text{d}t} = \mathbf{V}^T \mathbf{P} \mathbf{f}(\mathbf{q}, t).
\end{equation}

\noindent Scaling Eq. \ref{eq:fom} is essential to ensure that all equations contribute equally to the reduced system after projection. Without this, the reduced system could be biased by equations with larger magnitudes (e.g., the energy equation), leading to amplified floating-point errors. Using the low-rank representation from Eq. \ref{std. main rom} and substituting this into Eq. \ref{prj.rom}, a reduced-order ordinary differential equation (ODE) system can be obtained

\begin{equation}
\frac{\text{d} {\mathbf{q}_r}}{\text{d}t} = \mathbf{V}^T \mathbf{P} \mathbf{f}(\tilde{\mathbf{q}}, t), \quad {\mathbf{q_r}}(0) = \mathbf{V}^T \mathbf{q}_0,
\end{equation}

\noindent where the dimension of the ROM ODE is $n_p$, orders of magnitude smaller than $N$ in Eq \ref{eq:fom} (i.e, $n_p<<N$).

\subsection{Hyper-Reduction}
\label{sec:hyper-reduction}

Even though the projection-based MOR methods can produce a robust ROM with significantly lower dimensionality (\(n_p \ll N\)), evaluating the non-linear terms remains a bottleneck due to their computational complexity, which involves \(O(N)\) operations and becomes expensive for large-scale problems. A common solution to this problem is to use hyper-reduction to reduce the computational cost of evaluating these nonlinear terms by approximating them through a set of carefully selected sparse samples. Popular hyper-reduction methods include the discrete empirical interpolation method (DEIM) and gappy POD. The DEIM \cite{chaturantabut_nonlinear_2010} evaluates nonlinear terms at a small subset of components (i.e., sampled) and approximates the full-field non-linear terms via \emph{interpolation} in low-dimensional subspaces. On the other hand, gappy POD \cite{everson_karhunenloeve_1995} follows similar ideas but introduces oversampling in empirical interpolation and thus approximates the full-field nonlinear terms via a least-square \emph{regression} instead. The importance of oversampling has been discussed by several researchers \cite{peherstorfer_stability_2020,wentland_scalable_2023} and therefore, in the current study, we focus on using gappy POD to achieve hyper-reduction via the following formulation 

\begin{equation}
\mathbf{\tilde{f}} \approx \mathbf{U} (\mathbf{S}^T \mathbf{U})^{\dagger} \mathbf{S}^T \mathbf{f},
\end{equation}

\noindent where $(\mathbf{S}^T \mathbf{U})^{\dagger}$ denotes the Moore–Penrose pseudo-inverse of $\mathbf{S}^T \mathbf{U}$. The term \(\mathbf{S} \in \mathbb{R}^{N \times n_s}\) is a selection operator, belonging to a class of matrices with \(n_s\) columns (i.e, sampling points) of the identity matrix \(\mathbf{I} \in \mathbb{I}^{N \times N}\) and for oversampling via gappy POD, $n_s > n_p$. \(\mathbf{U} \in \mathbb{R}^{N \times n_d}\) is the basis set used to approximate the nonlinear term \(\mathbf{f}\). Typically, \(\mathbf{U}\) is constructed using POD from snapshots of \(\mathbf{f}\). It is found that setting \(\mathbf{U}\) to the trial POD basis \(\mathbf{V}\) (as in Eq. \ref{std. main rom}) also provides excellent approximations, which is adopted for all results presented in this paper. In addition, the empirical interpolation methods can be applied to obtain an approximation of the full-field FOM state, $\hat{\mathbf{q}} \in \mathbb{R}^N$, based on sparsely sampled local state information,

\begin{equation}
\hat{\mathbf{q}} = \mathbf{V}(\mathbf{S}^T \mathbf{V})^{\dagger} \mathbf{S}^T\mathbf{q}.
\label{deim}
\end{equation}

\section{Adaptive Model Order Reduction}
\label{sec:AROM}

As highlighted in previous sections, classical MOR methods often exhibit limited predictive capabilities due to the use of linear static bases, especially for convection-dominated problems with slow decay of Kolmogorov N-width \cite{bonnaillie-noel_efficient_2016,cohen_optimal_2020}. Several approaches have been proposed to lift this limitation through \emph{either} nonlinear bases \cite{lee_model_2020,kim_fast_2022,geelen_operator_2023,barnett_quadratic_2022} \emph{or} online adaptation techniques \cite{singh_lookahead_2023,huang_predictive_2023}. In the current work, we primarily focus on the online adaptation methods leveraging the work by Huang and Duraisamy \cite{huang_predictive_2023}, which has shown promise of incorporating adaptivity to construct truly predictive ROMs in chaotic, convection-dominated fluid flow problems. Similar to Huang and Duraisamy's work \cite{huang_predictive_2023}, we formulate an ideal optimization problem to update the basis, $\mathbf{V}$ and sampling points $\mathbf{S}$, simultaneously during the online ROM calculations,

\begin{equation}
\{\mathbf{V}^n,\mathbf{S}^n\} = \argmin_{\mathbf{V}^n \in \mathbb{R}^{N \times n_p} , \mathbf{S}^n \in \mathbb{R}^{N \times n_s}} \| \mathbf{V}^n(\mathbf{S}_n^T\mathbf{V}^n)^{\dagger}\mathbf{S}_n^T\mathbf{P}\mathbf{r}(\hat{\mathbf{q}}^n) \|_2^2 ,
\label{arom_main}
\end{equation}

\noindent where $\mathbf{r}({\hat{\mathbf{q}}}^n) \in \mathbb{R}^{N}$ is the fully discrete FOM equation residual in the form of, 

\begin{equation}    \mathbf{r}(\hat{\mathbf{q}}^{n}) = \hat{\mathbf{q}}^{n} - \hat{\mathbf{q}}^{n-1} - \Delta t \sum_{i=1}^{s} \beta_i \textbf{f}(\hat{\mathbf{q}}^{(i)}, t^{(i)}),
\label{eq:fom_eq_res}
\end{equation}

\noindent with the FOM state at the sampled points ($\mathbf{S}_n^T \mathbf{\hat{q}}^{n}$) directly estimated using Eq. \ref{eq:fom_eq_res} \emph{while} the FOM state at the unsampled points ($\mathbf{S}_n^{*T} \mathbf{\hat{q}}^{n}$) are found using Eq. \ref{deim} - i.e., $\mathbf{S}_n^{*T} \mathbf{\hat{q}}^{n} =  \mathbf{S}_n^{*T} \mathbf{V}^n(\mathbf{S}_n^T \mathbf{V}^n)^{\dagger} \mathbf{S}_n^T \mathbf{\hat{q}}^n$. Starting from this point, we omit the notation for the intermediate steps $\hat{\mathbf{q}}^{(j)}$ in time integration, as they follow the same form as shown in Eq. \ref{eq:SSPRK_intem}, to avoid redundancy. Moreover, it should be noted that different from the work by Huang and Duraisamy \cite{huang_predictive_2023}, we do not seek to solve for the reduced states $\mathbf{q}_r$ in the optimization problem posed in Eq. \ref{arom_main} \emph{while} instead, we directly leverage the sparse sampling approximation (Eq. \ref{deim}) to obtain the full-field state, the details of which are provided in the following section.

In the current work, we refer to Eq. \ref{arom_main} as the adaptive ROM formulation, where the basis and sampling points are both updated on the fly during the calculations. However, directly solving Eq. \ref{arom_main} is computationally prohibitive and may be intractable. To address this, we adopt a decoupled approach where we seek to solve two sequential minimization problems to adapt the basis and sampling points following a predictor-corrector idea. In the following sections, we provide detailed descriptions of the adaptive ROM formulation and algorithm.

\subsection{Adaptation of the basis }
\label{direc_basis_section}

First, we seek to adapt the basis matrix at time step $n$ from $\mathbf{V}^{n-1}$ to $\mathbf{V}^{n}$ which is formulated as the solution to the optimization problem

\begin{equation}
\argmin_{\mathbf{V}^{n} \in \mathbb{R}^{N \times n_p}} \| \mathbf{V}^{n} (\mathbf{V}^{n-1})^{\dagger} \mathbf{\hat{Q}}^n-\mathbf{\hat{Q}}^n \|_F^2,
\label{eq:basis_optimz}
\end{equation}

\noindent where $\mathbf{\hat{Q}}^n \in \mathbb{R}^{N \times w}$ is a data matrix containing a collection of estimated full states, $\mathbf{\hat{q}}^{n}$, spanning $w$ time instances

\begin{equation}
\mathbf{\hat{Q}}_n = [\hat{\mathbf{q}}^{n-w+1}, \dots,\mathbf{\hat{q}}^{n-1},\mathbf{\hat{q}}^{n}] \in \mathbb{R}^{N \times w}.
\label{data_window}
\end{equation}

\noindent We remark that the selection and computation of the estimated full states in constructing the data matrix, $\mathbf{\hat{Q}}^n$, plays a crucial role for basis adaptation, which determines the predictive capabilities of the resulting adaptive ROMs. We follow a similar approach in Ref \cite{huang_predictive_2023}, to construct $\mathbf{\hat{Q}}^n$. The full states at the sampled points, $\mathbf{S}_{n-1}^T\mathbf{\hat{q}}^{n}$, are estimated every time step with a physical time step of $\Delta t$ such that the FOM equation residual in Eq. \ref{eq:fom_eq_res} goes to zero - i.e, $ \mathbf{S}_{n-1}^T \mathbf{r} ( \mathbf{\hat{q}}^{n} )= 0$, which leads to

\begin{equation}
\mathbf{S}_{n-1}^T\mathbf{\hat{q}}^{n} = \mathbf{S}_{n-1}^T\mathbf{\hat{q}}^{n-1} + \Delta t \sum_{i=1}^{s} \beta_i \mathbf{S}_{n-1}^T\textbf{f}( \mathbf{S}_{n-1} \mathbf{S}_{n-1}^T\mathbf{q}^{(i)} + \mathbf{S}_{n-1}^{*} \mathbf{S}_{n-1}^{*T}\mathbf{q}^{(i)}, t^{(i)} ).
\label{full_est_sampled}
\end{equation}

\noindent The full states at the unsampled points, $\mathbf{S}^{*T}_{n-1}\mathbf{\hat{q}}^{n}$, with $\mathbf{S}_n\mathbf{S}_n^T + \mathbf{S}_n^{*}\mathbf{S}_n^{*^T} = \textbf{I}$, is estimated via hyper-reduction based on the full states at the sampled points, $\mathbf{S}_{n-1}^T\mathbf{\hat{q}}^{n}$, from Eq. \ref{full_est_sampled}

\begin{equation}
\mathbf{S}^{*T}_{n-1} \mathbf{\hat{q}}^{n} = \mathbf{S}^{*T}_{n-1} \mathbf{V}^{n}(\mathbf{S}_{n-1}^T \mathbf{V}^{n})^{\dagger}\mathbf{S}_{n-1}^T\mathbf{\hat{q}}^{n}. 
\label{estimate_deim}
\end{equation}



We remark that the full-state estimate strategies in Eqs. \ref{full_est_sampled} and \ref{estimate_deim} inherently assume 'local coherence' of the approximation in the solution domain \cite{peherstorfer_model_2020}. However, in a multi-physics problem such as combustion dynamics featuring both local coherence such as thin, disperse reaction regions and global coherence such as acoustics, this `local coherence' assumption is invalid. To address this issue, we follow the strategy in Ref \cite{huang_predictive_2023} to incorporate non-local effects by periodically estimating the full states at the unsampled locations infrequently by evaluating the full states every $z_s$ time steps ($1 \leq z_S$) with a larger time step, $z_s\Delta t$

\begin{equation}
    \mathbf{\hat{q}}^{n+z_s} = \mathbf{\hat{q}}^{n-1} + z_s\Delta t \sum_{i=1}^{s} \beta_i \textbf{f}(\mathbf{\hat{q}}^{(i)}, t^{(i)}).
\label{catch_up}
\end{equation}

\noindent However, it should be pointed out that we seek to evaluate the future full states in $z_s$ time steps (i.e., $\mathbf{\hat{q}}^{n+z_s}$) using Eq. \ref{catch_up} \emph{while} Ref \cite{huang_predictive_2023} seeks to re-evaluate the current full states (i.e., $\mathbf{\hat{q}}^{n}$) based on the past full states $z_s$ time steps back (i.e., $\mathbf{\hat{q}}^{n-z_s-1}$). We remark that this revision plays a crucial role in the formulation of the feature-guided sampling strategy in the next section.






Several methods have been proposed in the literature to solve the minimization problem presented in Eq. \ref{eq:basis_optimz} \cite{cortinovis_quasi-optimal_2020, peherstorfer_model_2020, peherstorfer_online_2015, zimmermann_geometric_2018, zucatti_adaptive_2023}. In the present work, we seek an exact solution to the minimization problem, which we refer to it as the direct basis adaptation method, 

\begin{equation}
\mathbf{V}^{n} = \mathbf{\hat{Q}}^n(\mathbf{V}^{n-1^\dagger}\mathbf{\hat{Q}}^n)^\dagger.
\label{dire_adapt}
\end{equation}

\noindent It can be easily shown that Eq. \ref{dire_adapt} represents one possible solution that satisfies the minimization problem introduced in Eq. \ref{eq:basis_optimz}. In addition, we remark that orthogonalizing the basis $\mathbf{V}$ after every basis adaptation to guarantee maintaining orthonormality and prevent any discrepancy growth in the newly generated basis. It is also important to keep in mind that all of the snapshots that are incorporated into data window matrix $\mathbf{Q}^n$ must be centered and normalized using $\mathbf{q}_{\text{ref}}$ and $\mathbf{P}$ respectively.

\subsection{Adaptation of the sampling points}
\label{subsec:adaptive_sampling}

Second, with the basis updated based on Eq. \ref{eq:basis_optimz}, we then seek to update the sampling points (selection operator) from $\mathbf{S}_{n-1}$ to $\mathbf{S}_n$, adopting similar ideas in Refs. \cite{uy_reduced_2022,peherstorfer_model_2020}, with the goal of minimizing the interpolation error arising from the gappy POD regression in approximating the estimated full states, $\mathbf{\hat{q}}^{n}$ from Eq. \ref{estimate_deim},

\begin{equation}
\mathbf{S}_n = \underset{\mathbf{S}_n \in \mathbf{S}^{N_s}}{\arg\min} \| \mathbf{\hat{q}}^{n} - \mathbf{\bar{\hat{q}}}^{n} \|_2^2,
\label{eq:interp_error}
\end{equation}

\noindent where $\mathbf{\bar{\hat{q}}}^{n} = \mathbf{V}^n ( \mathbf{S}_n^T \mathbf{V}^n )^{\dagger} \mathbf{S}_n^T \mathbf{\hat{q}}^{n}$ in which $\mathbf{V}^n$ represents the updated basis from Eq. \ref{eq:basis_optimz}. It has been proved by \cite{peherstorfer_stability_2020} that the error raised by sparse sampling (i.e, Eq. \ref{deim}) is bounded by,

\begin{equation}
\|\mathbf{q}-\mathbf{\hat{q}} \|_2 \leq \|(\mathbf{S}^T \mathbf{V})^{\dagger}\|^2_2 \|[\mathbf{I}-\mathbf{V}\mathbf{V}^T]\mathbf{q}\|^2_2,
\label{eq:interp_error_bound}
\end{equation}

\noindent where the first norm, $\|(\mathbf{S}^T \mathbf{V})^{\dagger}\|^2_2$, measures the contributions from the sampling errors and the second norm, $\|[\mathbf{I}-\mathbf{V}\mathbf{V}^T]\mathbf{q}\|^2_2$, quantities the contributions from the projection errors.

However, we remark that gappy POD introduces oversampling (i.e., $n_s > n_p$) to globally minimize the interpolation error in Eq. \ref{eq:interp_error} leads to a computationally intractable problem. Various methods have been proposed to select the sampling points \cite{carlberg_gnat_2013,peherstorfer_model_2020,wentland_scalable_2023} and most of them are greedy methods, that iteratively select points based on some measure of
local optimality. Therefore, improved greedy sampling methods are an area of
active research. In addition, though some work has been done to compare the performance of different sampling methods in conventional \emph{static} ROM construction \cite{peherstorfer_model_2020,wentland_scalable_2023}, to our best knowledge, very few work has been performed for \emph{adaptive} ROM investigations. Therefore, in the current work, we seek to extend the comparisons of different sampling methods to the adaptive ROM framework focusing on challenging convection-dominated problems with shocks and flames. Specifically, we select three methods from the literature, which include random sampling, eigenvector-based sampling \cite{peherstorfer_stability_2020}, and the Gauss–Newton with approximated
tensors (GNAT) \cite{carlberg_gnat_2013}. More importantly, we propose a feature-guided sampling method that is specifically designed for the adaptive ROM framework. For consistency, all sampling methods select the first $n_p$ sampling points using the QDEIM procedure by Drmac and Gugercin \cite{drmac_new_2015}, while the remaining $n_s - n_p$ points are selected by the aforementioned four methods. Here we describe each method:

\begin{enumerate}
    \item \emph{Random} sampling method simply places the $n_s - n_p$ oversampling points based on a uniform random distribution across the entire computational domain of interest. It requires no complex computations and is easy to implement for large-scale problems. In addition, it has been successfully demonstrated in constructing ROM for reacting flow problems \cite{huang_model_2022}. However, as shown by Wentland et al. \cite{wentland_scalable_2023}, random selection of the sampling points can lead to poor results.

    \item \emph{Eigenvector-based} method sampling seeks to minimize the sampling error term in Eq. \ref{eq:interp_error_bound} via a greedy approach. This method is a simplification by Peherstorfer et al. \cite{peherstorfer_stability_2020} of a greedy sampling procedure by Zimmermann and Willcox \cite{zimmermann_accelerated_2016} (Algorithm 4). To select the remaining \(n_s - n_p\) points, the method leverages the fact that by nature of the \(\ell^2\)-norm, the sampling error may be rewritten as,

    \begin{equation}
        \|(\mathbf{S}^T \mathbf{V})^\dagger\|_2 = \sigma_{\max} (\mathbf{S}^T \mathbf{V})^\dagger = \frac{1}{\sigma_{\min} (\mathbf{S}^T \mathbf{V})},
    \end{equation}
    
    where \(\sigma_{\max}\) and \(\sigma_{\min}\) indicate the maximum and minimum singular values of their arguments, respectively. Thus, at the \(m\)th sampling iteration, the row of \(\mathbf{V}\) is selected which maximizes the smallest singular value of \(\mathbf{S}^T_m \mathbf{V} \in \mathbb{R}^{n_s \times n_p}\). Here, \(\mathbf{S}^T_m \in \mathbb{R}^{n_s \times N}\) is the selection matrix constructed from the \(n_s\) unit vectors selected up to iteration \(m\) by the greedy procedure. It is shown in \cite{peherstorfer_stability_2020} that the bounds on this update of the smallest eigenvalue \(\lambda_{n_p}\), from update step \(m\) to \(m+1\), is given by
    
    \begin{equation}
        \lambda_{n_p}^{m+1} \geq \lambda_{n_p}^{m} + \frac{1}{2} \left( g + \|\mathbf{u}_+'\|_2^2 - \sqrt{(g + \|\mathbf{u}_+'\|_2^2)^2 - 4g ([\mathbf{z}_{n_p}^{m}]^T \mathbf{u}_+')^2} \right),
    \end{equation}
    
    where \(g = \lambda_{n_p-1}^{m} - \lambda_{n_p}^{m}\), and $\mathbf{z}_{n_p}^{m} \in \mathbb{R}^{n_p}$ is the eigenvector associated with eigenvalue \(\lambda_{n_p}^{m}\) (simply the \(n_p\)th canonical unit vector here). Here, \(\mathbf{u}_+' = \mathbf{Y}_m^T \mathbf{u}_{+}^T \in \mathbb{R}^{n_p}\) is defined for simplification, where \(\mathbf{Y}_m\) are the right singular vectors of \(\mathbf{S}^T_m \mathbf{V}\), and \(\mathbf{u}_+ \in \mathbb{R}^{n_p}\) is the row of \(\mathbf{V}\) to be added to the selection matrix iteration \(m\).

    \item \emph{GNAT} sampling, introduced by Carlberg et al. \cite{carlberg_gnat_2013}, builds upon the greedy sampling method initially developed for DEIM \cite{chaturantabut_nonlinear_2010}. This strategy aims to minimize errors in the regression of the basis vectors \( \mathbf{V}_m \) using a greedy approach, at each sampling step \( m \), the regression error is computed as  

    \begin{equation}
    \label{eq:GNAT}
    \mathbf{\epsilon}_m  = \mathbf{V}_m - \mathbf{V}_j \left(\mathbf{S}^T_m \mathbf{V}_j\right)^\dagger \mathbf{S}^T_m \mathbf{V}_m,
    \end{equation}
    
    where \( \mathbf{V}_j \in \mathbb{R}^{N \times j} \) consists of the first \( j \) basis vectors of \( \mathbf{V} \). The index \( j \) starts at 1 and increases by one in each sampling iteration until \( j \) reaches \( n_p \). At each step, the degrees of freedom corresponding to the largest absolute error \( |\mathbf{\epsilon}_m| \) are selected, and their associated canonical unit vectors are added to \( \mathbf{S}^T_m \). This method has been effectively employed in various aerodynamic applications \cite{carlberg_gnat_2013,carlberg_galerkin_2017}.

    \item \emph{Feature-guided sampling} (FGS) method directly leverages the adaptive ROM framework and places the sampling points at the regions corresponding to the highest gradients of representative features, $|\nabla \mathbf{\uptheta}|$, approximated based on the estimated future full states 

    \begin{equation}
    \mathbf{S}_n = \underset{\mathbf{S}_n \in \mathbf{S}^{n_s}}{\arg\max} \| \nabla \mathbf{\uptheta}^n \|_2^2,
    \label{eq:fgs_def}
    \end{equation}

    where $\mathbf{\uptheta}^n = \mathbf{\Theta}(\mathbf{\hat{q}}^{n+z_s})$ and $\mathbf{\Theta}: \mathbb{R}^N\to \mathbb{R}^N$ represents a nonlinear function that computes the target features based on $\mathbf{\hat{q}}^{n+z_s}$, the estimated future full states from Eq. \ref{catch_up}. It can be easily shown that by placing the sampling points in high-gradient regions, it inherently covers the rows of the basis, $\mathbf{V}$, exhibiting high gradients, leads to better condition numbers of $(\mathbf{S}^T \mathbf{V})$, and therefore reduces the sampling error, $\|(\mathbf{S}^T \mathbf{V})^{\dagger}\|^2_2$ in Eq. \ref{eq:interp_error_bound}. We remark that the proposed feature-guided sampling method shares similar motivation as the eigenvector-based approach \cite{peherstorfer_stability_2020} but resorts to a more direct physics-based strategy thanks to sampling based on the features evaluated from the future full states, $\mathbf{\hat{q}}^{n+z_s}$. It should also be noted that selecting appropriate features, $\mathbf{\uptheta}$ plays a crucial role and requires \emph{a priori} physical knowledge of the dynamics in the target problems but we also remark that such knowledge is commonly physically intuitive and easy to impose. For example, for shock-dominated problems, pressure ($p$) well represent the dominant crucial physics, while for flame-dominated problems, temperature ($T$) serves as a better choice for feature. 
    
\end{enumerate}

\begin{figure}[ht]
  \centering
    \includegraphics[width=0.8\linewidth]{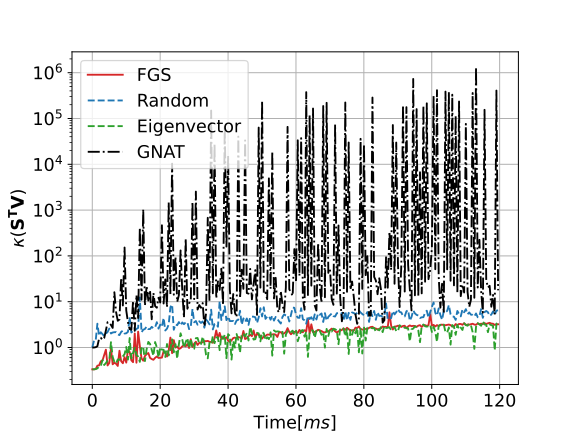}
  \caption{Comparison of the condition number ($\mathbf{S}^T \mathbf{V}$) using different sampling methods.}
  \label{fig:sod_shock_cond} 
\end{figure}

To demonstrate the effects of four different sampling methods on the sampling errors, $\|(\mathbf{S}^T \mathbf{V})^{\dagger}\|^2_2$ in Eq. \ref{eq:interp_error_bound}, we perform an \emph{offline} synthetic analysis on the snapshots data from a Sod shock tube simulation (details of the simulation is included in section \ref{sec:sod_shock}) by comparing the condition number of the sampled basis matrix, $\kappa(\mathbf{S}^T \mathbf{V})$, an explicit measure of the sampling errors. Specifically, the basis $\mathbf{V}$ is updated using Eq. \ref{dire_adapt} by directly using the FOM snapshots to construct the data matrix, $\mathbf{\hat{Q}}$, so that the projection errors, $\|[\mathbf{I}-\mathbf{V}\mathbf{V}^T]\mathbf{q}\|^2_2$, in Eq. \ref{eq:interp_error_bound} remain low. The comparisons are shown in Fig. \ref{fig:sod_shock_cond} with 5 POD modes included ($n_p = 5$), total 10 sampling points selected ($n_s = 10$) and for consistency between different methods, first 5 selected using the QDEIM procedure, while the remaining 5 are chosen based on the respective oversampling strategies discussed above. As shown in Fig. \ref{fig:sod_shock_cond}, the FGS method (red solid line) exhibits significantly lower condition numbers and reduced oscillations compared to the GNAT (black dash-dot line) and Random (blue dashed line) methods, which potentially leads to a more stable, well-conditioned system and more importantly lower sampling error. In addition,  the eigenvector-based method (green dashed line) shows comparable performance as the FGS but we remark that the FGS method is more computationally efficient than the eigenvector-based method. While the eigenvector approach relies on costly SVDs or eigenvalue computations, FGS achieves a similar performance at a much lower computational cost. We provide a detailed comparisons of the computational complexity between FGS and eigenvector-based methods in section \ref{subsec:flops}.This makes FGS a highly effective and practical choice for large-scale problems. The influence of this sampling technique on both the condition number and interpolation error is detailed in the results for the Sod shock tube case in section \ref{sec:results}.


Finally, we remark that the adaptation of the sampling points requires evaluating the full states, $\mathbf{\hat{q}}^{n+z_s}$, at all the points, which incurs high computational costs that scale with the total number of degrees of freedom, $N$, in the FOM. Therefore, following the work in Refs. \cite{huang_predictive_2023,peherstorfer_model_2020}, we choose to adapt the sampling points every $z_s$ time steps, which is consistent with the non-local full-state estimate in Eq. \ref{catch_up} and mitigates the penalty of additional expensive evaluations of the full states. It is evident that \(z_s\) is a crucial parameter in both the adaptation of the basis and the sampling points, affecting the accuracy and efficiency of the adaptive ROM as shown by Huang and Duraisamy \cite{huang_predictive_2023}. The details of the feature-guided adaptive ROM algorithm is provided in section \ref{sec:algorithm}.


\section{Computational Algorithm and Complexity}
\label{sec:algorithm}

In this section, we describe the computational procedure and complexity of constructing the feature-guided adaptive ROM for practical considerations.

\subsection{Computational algorithm}
\label{subsec:algorithm}

The proposed feature-guided adaptive ROM algorithm is summarized Algorithm \ref{AROM_Algorithm}, which takes the following inputs:
\begin{enumerate}
    \item $M$: total number of physical time steps;
    \item $w_{init}$: initial training window size;
    \item $z_s$: frequency of updating sample points;
    \item $n_s$: the total number of sampling points.
\end{enumerate}

The adaptive ROM is initialized by solving the FOM over a small time window of $w_{init}$ time steps. The collected snapshots are used to form the initial data matrix, $\mathbf{\hat{Q}}^0$ (line 2). Next, the reference state $\mathbf{q}_{\text{ref}}$ and scaling matrix $\mathbf{P}$ are determined based on $\mathbf{\hat{Q}}^0$ following the procedures in section \ref{subsec:std_MOR} (line 3). This data matrix is used to compute initial basis $\mathbf{V}^0$ matrix by performing SVD in Eq. \ref{eq:svd} (line 4). The algorithm then continues to advance the online adaptive ROM computation from time step $w_{init}+1$ to $M$.

For the first time step after the initial window (i.e., $n =w_{init}+1$), the future full state, $\mathbf{q}^{n+z_s}$, is estimated using Eq. \ref{catch_up} (line 8). Next, the full states at the sampled ($\mathbf{S}_{n-1}^T\mathbf{\hat{q}}^n$) and unsampled ($\mathbf{S}_{n-1}^{*T}\mathbf{\hat{q}}^n$) points are estimated using Eqs. \ref{full_est_sampled} and \ref{estimate_deim} respectively (line 10), which are used to construct the full-field full states ($\mathbf{\hat{q}}^n$) (line 12) and update in data matrix $\mathbf{\hat{Q}}^n$ following Eq. \ref{data_window} (line 15). Consequently, the trial basis, $\mathbf{V}^n$, is updated using Eq. \ref{dire_adapt}  (line 14) and the sampling points, $\mathbf{S}_n$, are updated using feature-guided algorithm summarized in Algorithm \ref{Sampling Algorithm} (line 15). Then advancing from time step $w_{init}+2$ to $M$, for every $z_s$ time steps, the future full state, $\mathbf{q}^{n+z_s}$, is estimated using Eq. \ref{catch_up} (line 17) with the the full states at the sampled points ($\mathbf{S}_{n-1}^T\mathbf{\hat{q}}^n$) esimated using \ref{full_est_sampled} (line 19) \emph{while} otherwise the full states at the sampled ($\mathbf{S}_{n-1}^T\mathbf{\hat{q}}^n$) and unsampled ($\mathbf{S}_{n-1}^{*T}\mathbf{\hat{q}}^n$) points are estimated using Eqs. \ref{full_est_sampled} and \ref{estimate_deim} respectively (line 27). The collected full states, $\mathbf{\hat{q}}^n$, are used to update the data matrix, $\mathbf{\hat{Q}}^n$, (lines 20 and 28), which are then used to update the trial basis, $\mathbf{V}^n$ (lines 21 and 29). In addition, for every $z_s$ time steps, the sampling points, $\mathbf{S}_n$, are updated using feature-guided algorithm (line 22) \emph{while} otherwise the sampling points remain unchanged.

\begin{algorithm}
\caption{Adaptive ROM algorithm}
\label{AROM_Algorithm}
\begin{algorithmic}[1]

\State \textbf{Input}: $M, w_{init}, z_s, n_s$

\State Solve FOM (Eq. \ref{eq:fom}) for $w_{init}$ time steps to collect the initial data matrix $\hat{\mathbf{Q}}^0$

\State Compute the reference state, $\mathbf{q}_{\text{ref}}$ and the scaling matrix $\mathbf{P}$ as explained in section \ref{subsec:std_MOR}

\State Compute the initial basis, \textbf{$\mathbf{V}^{0}$} using Eq. \ref{eq:svd}

\State Compute the initial sampling points, \textbf{$\mathbf{S}_{0}$} 
\For{$n = w_{init}+1 \to M$}

\If{$n == w_{init} + 1$}

\State Calculate \text{$\mathbf{{q}}^{n+z_s}$} using Eq. \ref{catch_up}

\State Estimate the full states at the \emph{sampled} and \emph{unsampled }points: 
\State \ \ \ \ \ \ \ \textbf{$\mathbf{S}_{n-1}^T \mathbf{\hat{q}}^{n}$} using Eq. \ref{full_est_sampled} and $\mathbf{S}^{*T}_{n-1} \mathbf{\hat{q}}^{n}$ using Eq. \ref{estimate_deim}

\State Construct the full-field full states: 
\State \ \ \ \ \ \ \ $\mathbf{\hat{q}}^n=\mathbf{S}_{n-1}\mathbf{S}_{n-1}^T\mathbf{\hat{q}}^n + \mathbf{S}^{*}_{n-1}\mathbf{S}_{n-1}^{*T}\mathbf{\hat{q}}^n$

\State Update the data matrix \textbf{$\hat{\mathbf{Q}}^n$} following Eq. \ref{data_window}

\State Update the trial basis using Eq. \ref{dire_adapt} $\mathbf{V}^{n-1} \rightarrow \mathbf{V}^{n}$

\State Update sampling points \textbf{$\mathbf{S}_{n-1} \rightarrow \mathbf{S}_{n}$} following Algorithm \ref{Sampling Algorithm}

\ElsIf{$mod(n,z_s)==0$}

\State Calculate \text{$\mathbf{{q}}^{n+z_s}$} using Eq. \ref{catch_up}

\State Estimate the full states at the \emph{sampled} points: 
\State \ \ \ \ \ \ \ \textbf{$\mathbf{S}_{n-1}^T \mathbf{\hat{q}}^{n}$} using Eq. \ref{full_est_sampled}

\State Update the data matrix \textbf{$\hat{\mathbf{Q}}^n$} following Eq. \ref{data_window}

\State Update the trial basis using Eq. \ref{dire_adapt} $\mathbf{V}^{n-1} \rightarrow \mathbf{V}^{n}$

\State Update sampling points $\mathbf{S}_{n-1} \rightarrow \mathbf{S}_{n}$ following Algorithm \ref{Sampling Algorithm}

\Else

\State Estimate the full states at the \emph{sampled} and \emph{unsampled }points: 
\State \ \ \ \ \ \ \ \textbf{$\mathbf{S}_{n-1}^T \mathbf{\hat{q}}^{n}$} using Eq. \ref{full_est_sampled} and \textbf{$\mathbf{S}^{*T}_{n-1} \mathbf{\hat{q}}^{n}$} using Eq. \ref{estimate_deim}

\State Construct the full-field full states: 
\State \ \ \ \ \ \ \ $\mathbf{\hat{q}}^n=\mathbf{S}_{n-1}\mathbf{S}_{n-1}^T\mathbf{\hat{q}}^n + \mathbf{S}^{*}_{n-1}\mathbf{S}_{n-1}^{*T}\mathbf{\hat{q}}^n$

\State Update the data matrix \textbf{$\hat{\mathbf{Q}}^n$} following Eq. \ref{data_window}

\State Update the trial basis using Eq. \ref{dire_adapt} \textbf{$\mathbf{V}^{n-1} \rightarrow \mathbf{V}^{n}$}

\EndIf

\EndFor
\end{algorithmic}
\end{algorithm}

Algorithm \ref{Sampling Algorithm} outlines the feature-guided sampling (FGS) method to update the sampling points, which requires the following inputs: 
\begin{enumerate}
    \item $\mathbf{\hat{q}}^{n+z_s}$: the estimated future full state from Algorithm \ref{AROM_Algorithm};

    \item $n_s$: the total number of sampling points; 

    \item $\mathbf{V}^{n}$: the updated trial basis at the current time step;

    \item $\uptheta$: the target feature.
\end{enumerate}

The FGS method determines the first $n_p$ (same as the dimension of the trial basis, $\mathbf{V}^{n}$) sampling points using QDEIM \cite{drmac_new_2015}, denoted as $\mathbf{\bar{S}}_n^T$ (line 2). If the specified number of sampling points is lower than $n_p$ (i.e., $n_s \leq n_p$), the first $n_p$ samples in $\mathbf{\bar{S}}_n^T$ are taken (line 4); otherwise, the magnitudes of the gradients of the selected features, $\| \nabla \mathbf{\uptheta}^n \|_2^2$, are computed with $\mathbf{\uptheta}^n = \mathbf{\Theta}(\mathbf{\hat{q}}^{n+z_s})$ as defined in Eq. \ref{eq:fgs_def} (line 7). The rest of the sampling points are then placed on the $n_{s} -n_{p}$ points corresponding to the largest magnitudes of $\mathbf{e}^n_{\mathbf{\uptheta}}$ (line 9-10), denoted as $\mathbf{\hat{S}}_n^T$. Finally, the resulting sampling points, $\mathbf{S}_n^T$, are constructed as a combination of $\mathbf{\bar{S}}_n^T$ and $\mathbf{\hat{S}}_n^T$ (line 11).

\begin{algorithm}
\caption{Feature-guided sampling (FGS) algorithm}
\label{Sampling Algorithm}
\begin{algorithmic}[1]
\State \textbf{Input}:  $\mathbf{\hat{q}}^{n+z_s}$, $n_s$, $\mathbf{V}^{n}$, $\uptheta$
\State Determine $n_p$ sampling points $\mathbf{\bar{S}}_n^T$ using QDEIM \cite{drmac_new_2015}

\If{$n_s \leq n_p$}

\State $\mathbf{S}_n^T = \mathbf{\bar{S}}_n^T$($n_p$,:)

\Else

\State Compute the magnitudes of the gradients of the selected features, $\mathbf{e}^n_{\mathbf{\uptheta}}$:
\State \ \ \ \ \ \ \ $\mathbf{e}^n_{\mathbf{\uptheta}} =\| \nabla \mathbf{\uptheta}^n \|_2^2$ with $\mathbf{\uptheta}^n = \mathbf{\Theta}(\mathbf{\hat{q}}^{n+z_s})$ as defined in Eq. \ref{eq:fgs_def}

\State Place the remaining $n_s-n_p$ sampling points:
\State \ \ \ \ \ \ \ $[\sim , \hat{\mathbf{i}}]$ = \texttt{sort}($|\mathbf{e}^n_{\mathbf{\uptheta}}|$,`descend')
\State \ \ \ \ \ \ \ $\hat{\mathbf{i}}[1:(n_s-n_p)]$ is selected for $\mathbf{\hat{S}}_n^T$

\State Construct the sampling points: $\mathbf{S}_n\mathbf{S}_n^T = \mathbf{\bar{S}}_n\mathbf{\bar{S}}_n^T$ + $\mathbf{\hat{S}}_n\mathbf{\hat{S}}_n^T$

\EndIf
\end{algorithmic}
\end{algorithm}

\subsection{Computational complexity}
\label{subsec:flops}

Next, we present a detailed analysis on the computational complexity of the featured-guided adaptive ROM algorithm based on the floating-point operations (FLOPs). A FLOP refers to either a floating-point addition or multiplication, with no distinction in computational cost. Our analysis focuses on the complexity of both FOM and ROM using explicit four-stage SSP-RK3 integration scheme.

First, we estimate the FLOPs required for FOM calculations, denoted as $\text{FLOP}_\text{FOM}$, for one physical time step as summarized in Table \ref{tab:FOM_FLOPS}.

\begin{table}[ht]
    \centering
    \caption{Approximated FLOPs for FOM calculation for one time step}
    \footnotesize 
    \begin{tabular}{lccc}
        \toprule
         \textbf{Operations} & \textbf{Approximated FLOPs} \\
        \midrule
        Evaluate the non-linear FOM equation residual $\mathbf{f}(\mathbf{q}^n)$ & $N_{var}^2N_{elem}$ \\
        Four-stage SSP-RK3 Time Integration & $4N_{var}^2N_{elem}$ \\
        \midrule
        Total & $5N_{var}^2N_{elem}$ \\
        \bottomrule
    \end{tabular}
    \label{tab:FOM_FLOPS}
\end{table}

Next, we estimate the FLOPs for the conventional static-basis hyper-reduced Galerkin ROM (referred to as \emph{static ROM} for brevity) in Table \ref{tab:ROM_FLOPS}, denoted as $\text{FLOP}_\text{SROM}$, over one physical time step to establish a baseline for comparison with the adaptive ROM.

\begin{table}[ht]
    \centering
    \caption{Approximated FLOPs for hyper-reduced Galerkin ROM calculation for one time step}
    \footnotesize 
    \begin{tabular}{lccc}
        \toprule
         \textbf{Operations} & \textbf{Approximated FLOPs} \\
        \midrule
        Evaluate residual only at sampled points $\mathbf{S}^T \mathbf{f}(\mathbf{q}^n)$ & $N_{var}n_s$ \\
        Evaluate the psudo-inverse $(\mathbf{S}^T \mathbf{V})^\dagger$ & $N_{var}n_sn_p^2$ \\
        Estimate full-rank residual $(\mathbf{S}^T \mathbf{V})^\dagger \mathbf{S}^T \mathbf{f}(\mathbf{q}^n)$ & $2N_{var}n_sn_p$ \\
        Four-stage SSP-RK3 Time Integration at low-rank& $4n_p$ \\
        \midrule
        Total & $n_sN_{var}(1+n_p^2+2n_p)+4n_p$ \\
        \bottomrule
    \end{tabular}
    \label{tab:ROM_FLOPS}
\end{table}

Finally, we estimate the FLOPs for the feature-guided adaptive ROM based on the computing procedures in Algorithm \ref{AROM_Algorithm}, denoted as $\text{FLOP}_\text{AROM}$, which are summarized in Table \ref{tab:AROM_FLOPS}. We remark that incorporating non-local information into the adaptive ROM does not introduce additional computational complexity, as the same update rate, $z_s$, applies to both sampling point selection and non-local full-state estimation, the latter of which requires full-state estimation at all points.

\begin{table}[ht]
    \centering
    \caption{Approximated FLOPs for feature-guided adaptive-ROM calculation in Algorithm \ref{AROM_Algorithm} for one time steps on average}
    \footnotesize 
    \begin{tabular}{lccc}
        \toprule
         \textbf{Operations} & \textbf{Approximated FLOPs} \\
        \midrule
        Propogate Sampled FOM $\mathbf{S}^T \mathbf{\hat{q}}^{n}$ & $5N_{var}^2n_s$ \\
        Estimate full-state $\mathbf{\hat{q}}^n$ & $N_{var}n_s(n_p^2+2n_p+1)$ \\
        Update Basis & $wn_p(w+2N_{var}n_s)$ \\
        Update Samples every $z_s$ & $\frac{(n_p+2)N_{var}N_{elem}+(n_p+1)n_s}{z_s}$ \\
        \midrule
        Total & $5N_{var}^2n_s + N_{var}n_s(n_p^2+2n_p+1)$\\  & $+ wn_p(w+2N_{var}n_s) + \frac{(n_p+2)N_{var}N_elem+(n_p+1)n_s}{z_s} $\\
        \bottomrule
    \end{tabular}
    \label{tab:AROM_FLOPS}
\end{table}

Furthermore, the ratio of FLOPs between FOM and ROM is defined to quantify the  efficiency gain of the ROM:

\begin{equation}
    \lambda = \frac{\text{FLOP}_\text{FOM}}{\text{FLOP}_\text{ROM}}. 
    \label{eq:ROM_efficiency_gain}
\end{equation}

\begin{figure}[ht]
  \centering
    \includegraphics[width=0.75\linewidth]{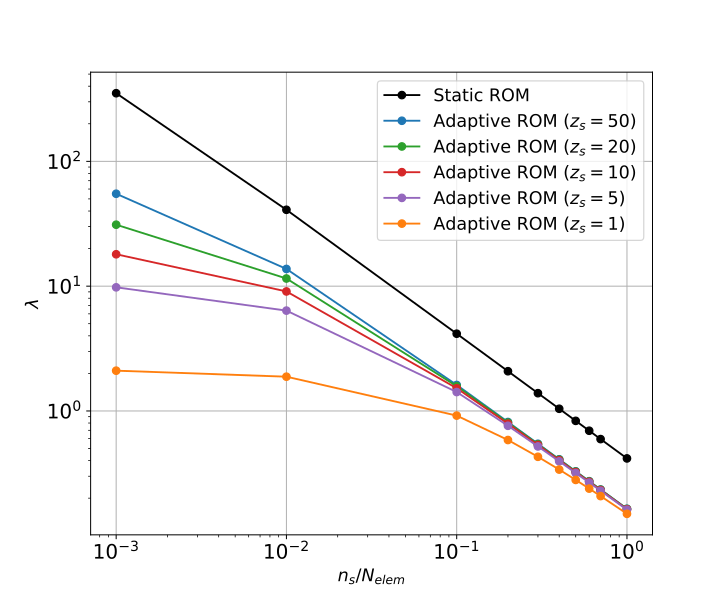} 
  \caption{Approximate efficiency gain of static and adaptive ROM (with different $z_s$ adopted) versus sampling rate}
  \label{fig:flops} 
\end{figure}

\noindent The theoretical efficiency gain of both static and adaptive ROMs are estimated and compared against the fraction of sampling points, \( n_s / N_{\text{elem}} \), using \( n_p = 5 \), \( w = 5 \), as shown in Fig. \ref{fig:flops}, with different \( z_s \) values for the adaptive ROM. It can be readily seen that for both static and adaptive ROMs, efficiency gains increase as the number of sampling points decreases. However, while the efficiency gain of the static ROM grows almost linearly with the reduction in sampling points, given the additional operations required to update the basis and sampling points, the adaptive ROM's efficiency gain follows an asymptotic trend and is not expected to achieve the same level of computational efficiency gains as the static ROM. In addition to
the number of sampling points, the rate at which sampling points are updated \( z_s \), determines the acceleration that the adaptive ROM can achieve.



In addition, we present and compare the FLOPs required for FGS and eigenvector-based method \cite{zimmermann_accelerated_2016,peherstorfer_stability_2020} in Table \ref{tab:FGS_vs_Eigen_FLOPS}. First, it should be noted that both the FGS and eigenvector-based sampling methods leverage the QDEIM technique to select the initial set of samples, which requires roughly $n_pN^2$ numerical operations. Second, as demonstrated in Table \ref{tab:FGS_vs_Eigen_FLOPS}, the remaining computations for both methods depend significantly on the problem size. However, the numerical complexity of the FGS method increases at a slower rate compared to the eigenvector-based method, which indicates that the eigenvector-based method incurs computational costs that scale more dramatically with problem size than the FGS. Detailed numerical complexity analysis is presented in \ref{app4}.

\begin{table}[ht]
    \centering
    \caption{Approximated FLOPs for different sampling methods}
    \begin{tabular}{lccc}
        \toprule
         \textbf{Method} & \textbf{Approximated FLOPs} \\
        \midrule
        FGS & $n_pN^2+2N_{elem}+N_{elem}log(N_{elem})$ \\
        Eigenvector-based & $n_pN^2+(n_p^3+Nlog(N))(N_{var}n_s-n_p)$ \\
        \bottomrule
    \end{tabular}
    \label{tab:FGS_vs_Eigen_FLOPS}
\end{table}

\section{Numerical Results and Analysis}
\label{sec:results}

To evaluate the performance and capabilities of the adaptive ROM method with the proposed feature-guided sampling strategies in modeling complex multi-physics problems, a suite of one-dimensional (1D) test problems are considered, which contain strong transient (i.e., non-stationary) convection of physics featuring \emph{either} sharp gradients \emph{or} discontinuities, resulting in slow decays of Kolmogorov N-width. The first test problem simulates the Sod shock tube \cite{sod_survey_1978}, a classic test for the accuracy of computational fluid codes, which incorporates many features challenging for ROM, such as advection of discontinuous physics (i.e., shocks) and multi-scale dynamics of subsonic wave expansions. The second test problem selects a benchmark 1D freely propagating premixed laminar flame that has been used to evaluate the performance of different ROM techniques \cite{huang_model_2022}. Specifically, it features strong advection of sharp gradients (i.e., the flame propagation), and nonlinear dynamics due to chemical reactions. The third test problem is a 1D detonation tube \cite{oran_weak_1982}, which simulates the formation of a detonation wave and involves challenging physics of both shocks and chemical reactions, present in the previous two test problems. The formation process of the detonation wave is well recognized to be difficult if not impossible for ROM development. We remark that though simulated with 1D models, the proposed test suite inherits many features that are challenging for ROM development, which include convection-dominated transport and multi-scale, multi-physics coupling. More importantly, the constructed 1D problems allow evaluations of ROM capabilities in great detail without incurring an exorbitant computational cost. An open-source 1D CFD code\footnote{\url{https://github.com/alimike97/ROMify_Public}} has been developed and is employed for both the FOM and ROMs, with the capability of simulating chemically reactive flows using an open-source package, Cantera \cite{speth_canteracantera_2017}. The code solves the conservation equations for mass, momentum, energy, and species transport in a fully coupled manner. Further details on the FOM equations are included in \ref{app1}. The FOM utilizes a cell-centered, second-order accurate finite volume method for spatial discretization. The Roe scheme \cite{roe_approximate_1981} is used for calculating inviscid fluxes. To maintain flow field monotonicity in the presence of strong gradients, a gradient limiter by Barth and Jespersen \cite{barth_design_1989} is employed, along with a ghost cell approach for boundary conditions. Time integration of all simulations is performed using an explicit four-stage third degree strong stability Runge-Kutta (four-stage SSP-RK3)\cite{durran_numerical_2010}. All the FOM and ROM calculations in the rest of this section are performed on one compute node (Dual Intel Xeon 6240R CPU) with 192GB2933 MHz DDR memory.

\subsection{Sod shock tube}
\label{sec:sod_shock}

\begin{figure}[ht]
  \centering
    \includegraphics[width=0.7\linewidth]{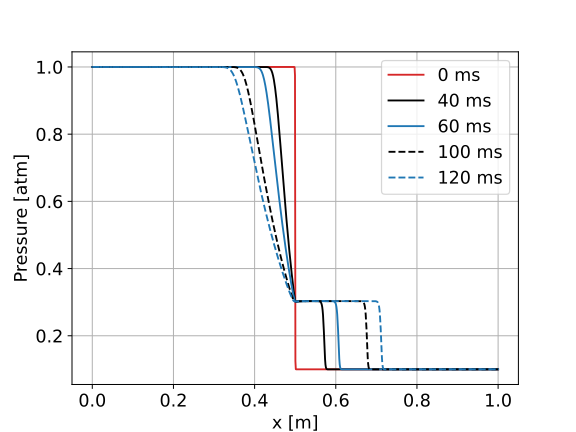} 
  \caption{Representative FOM solutions at different time instances for the Sod shock tube.}
  \label{fig:sod_shock_rep_FOM} 
\end{figure}

First, we consider the Sod shock tube test problem, the computational domain of which has a length of $1\, m$, discretized with 1000 uniform finite volume cells. Characteristic boundary conditions are specified at the two ends of the domain. The FOM calculation is initialized using a quiescent flow with a pressure discontinuity at the center of the domain (i.e., $x = 0.5 \, m$), corresponding to a pressure drop of a factor of 10, with $p = 1 \, atm$ on the left and $p = 0.1 \, atm$ on the right. The unsteady FOM solution is advanced from $0$ to $120\, ms$ using a constant physical time step size of $\Delta t = 50\, \mu s$.  Representative FOM solutions are shown in Fig. \ref{fig:sod_shock_rep_FOM} from $0\, ms$ to $120\, ms$. It can be readily seen that the high-pressure gas on the left expands rapidly into the lower pressure region on the right, leading to a series of non-linear waves such as shock wave that moves into the low-pressure gas, rapidly compressing and heating it along with an expansion fan wave (or rarefaction wave) that propagates back into the high-pressure gas, decreasing the pressure and density.


\subsubsection{POD characteristics}

The characteristics of the POD basis are investigated to access how well it can represent the dynamics in the FOM dataset, using the POD residual energy

\begin{equation}
\text{POD Residual Energy}(n_p), \% = \left(1 - \frac{\sum_{i=1}^{n_p} \tilde{\sigma}_i^2}{\sum_{i=1}^{n_{p,\text{total}}} \tilde{\sigma}_i^2}\right) \times 100
\label{eq:POD_res}
\end{equation}

\noindent where $\tilde{\sigma}_i$ is the $i^{th}$ singular value of the SVD used to compute the trial basis $\mathbf{V}$ and are arranged in descending order. $n_p$ is the number of modes retained in the POD trial basis, and $n_{p,total}$ is the total number of modes in the training dataset.

\begin{figure}[ht]
  \centering
    \includegraphics[width=0.8\linewidth]{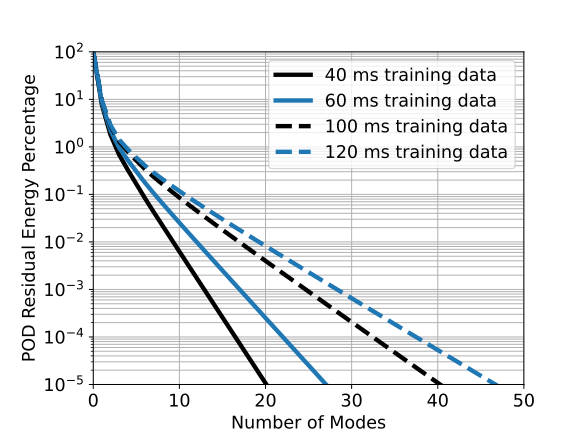} 
  \caption{POD residual energy distribution for the Sod shock tube with different training snapshots to compute the POD basis}
  \label{fig:sod_shock_pod} 
\end{figure}

\begin{figure}[ht]
  \centering
    \includegraphics[width=0.8\linewidth]{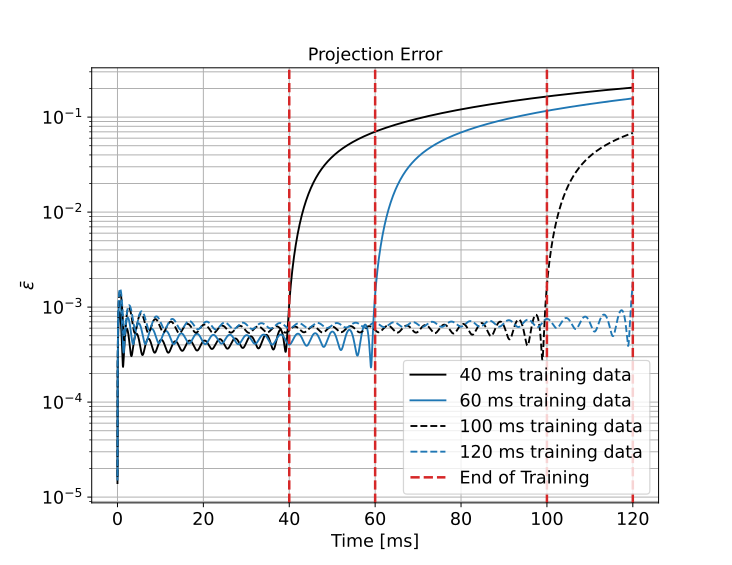} 
  \caption{Projection error for the Sod shock tube with different training snapshots to compute the POD basis.}
  \label{fig:sod_shock_proj_error} 
\end{figure}

The POD residual energy is investigated in Fig. \ref{fig:sod_shock_pod} by including different numbers of FOM snapshots from $t = 0$ to compute the trial basis, $\mathbf{V}$, and it can be seen that by increasing the number of training snapshots, the number of POD modes required to capture the same level of residual energy increases proportionally, leading to a much slower decay of the POD residual energy, and therefore, a slow decay of Kolmogorov N-width. For instance, to capture approximately $99.99999 \%$ of the total energy, 20 modes are required for training snapshots of $40 \, ms$, 27 modes for $60\, ms$, 40 modes for $100\, ms$, and 46 modes for $120 \, ms$. In addition, to further investigate the sufficiency of using the computed trial basis $\mathbf{V}$ to represent dynamics beyond the training snapshots, we quantify the projection error based on the normalized and variable-averaged \(\ell^2\) error 

\begin{equation}
\mathbf{\bar{\varepsilon}}^n = \frac{1}{N_{\text{var}}} \sum_{i=1}^{N_{\text{var}}} 
\frac{\left\| \mathbf{\bar{q}}_{i}^n - \mathbf{q}_{i}^n \right\|_2}{\left\| \mathbf{q}_{i}^n \right\|_2},
\label{eq:proj_err}
\end{equation}

\noindent where $\mathbf{\bar{q}}_{i}^n$ represents the $i^{\text{th}}$ solution variable of the state vector, $\mathbf{q}_{i}^n$, at time step $n$, evaluated as, $\mathbf{q}_{\text{ref}} + \mathbf{P}^{-1} \mathbf{V} \mathbf{V}^T \mathbf{q}^n$.

The projection errors are calculated using different training snapshots and are compared in Fig. \ref{fig:sod_shock_proj_error} by including the number of modes corresponding to $99.99999 \%$ of the total energy as shown in Fig. \ref{fig:sod_shock_pod}. It can be seen that the projection errors are low within the training region while they start to increase significantly and remain high outside the training region. This observation directly indicates the insufficiency of using these trial bases in representing dynamics beyond the training period, which implies poor predictive capabilities of the resulting ROMs based on these bases.

\subsubsection{Performance of adaptive ROM with different sampling strategies}

The adaptive ROMs are constructed using different sampling strategies discussed in section \ref{subsec:adaptive_sampling}, including (1) Feature-guided sampling (FGS), (2) Random, (3) Eigenvector-based, and (4) GNAT. For consistency, the ROM performance is evaluated based on the whole window for the testing dataset (0 to $120\, ms$). In this problem, we select the magnitude of pressure gradients as the feature for FGS. For quantitative evaluations of the adaptive ROMs, we implement a multi-level error measurement on ROM accuracy, which includes: 

\noindent (1) the normalized, time-averaged, and variable-averaged total \(\ell^2\) error between FOM and adaptive ROM solutions to access the overall accuracy of the constructed adaptive ROMs

\begin{equation}
   \epsilon_{total}^n =  \frac{1}{N_{\text{var}}} \sum_{i=1}^{N_{\text{var}}} \frac{\|\mathbf{q}_{FOM,i}^n - \mathbf{q}_{ROM,i}^n\|_2^2}{\|\mathbf{q}_{FOM,i}^n\|_2^2},
   \label{eq:total_err}
\end{equation}

\noindent where ${\mathbf{q}}_{FOM,i}^n$ and ${\mathbf{q}}_{ROM,i}^n$ represent the $i^{th}$ state variable solution of FOM and adaptive ROM at time step $n$;

\noindent (2) the projection \(\ell^2\) error defined in Eq. \ref{eq:proj_err} to evaluate the sufficiency of the updated basis in representing the target dynamics;

\noindent (3) and the interpolation \(\ell^2\) error to evaluate the effectiveness of the sampling-point selection in capturing the key dynamics in the problem

\begin{equation}
    \epsilon_{interp}^n = \frac{1}{N_{\text{var}}} \sum_{i=1}^{N_{\text{var}}} \frac{\| \mathbf{q}_{ROM,i}^n-\mathbf{V}^n(\mathbf{S}_n^T \mathbf{V}^n)^{\dagger}\mathbf{S}_n^T\mathbf{q}_{ROM,i}^n\|_2^2}{\| \mathbf{q}_{ROM,i}^n\|_2^2}.
    \label{eq:interp_err}
\end{equation}

Following Algorithm \ref{AROM_Algorithm}, the adaptive ROMs are initialized with $0.25 \, ms$ snapshots  ($w_{init} = 5$) from t = 0 and developed with 5 modes ($n_p = 5$ retrieving 100\% of the energy) and 10 sampling points ($n_s = 10$ covering 1\% of the entire computational domain), which are updated every 10 time steps ($z_s = 10$). For consistency, we remark that the first 5 sampling points in FGS, random, and eigenvector-based, are selected using QDEIM with the rest 5 points selected based on different oversampling strategies \emph{while} the 10 sampling points in GNAT are determined following the procedures described in section \ref{subsec:adaptive_sampling}.

\begin{figure}[ht]
  \centering
    \includegraphics[width=0.8\linewidth]{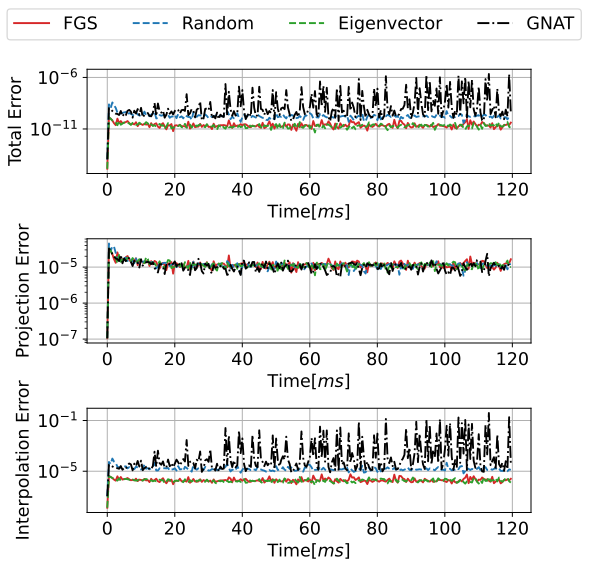} 
  \caption{Comparisons of the multi-level error measurement between the adaptive ROMs adopting different sampling methods for the Sod shock tube}
  \label{fig:sod_shock_error} 
\end{figure}

The multi-level error measurement is implemented and compared for the adaptive ROMs with different sampling methods in Fig. \ref{fig:sod_shock_error}, which shows that using FGS and eigenvector-based in adaptive ROM construction produces approximately one order of magnitude improvement in total errors (Fig. \ref{fig:sod_shock_error} top) over random and GNAT. With very similar projection errors between different sampling methods (Fig. \ref{fig:sod_shock_error} middle), the improvement in the total errors can be mainly attributed to the significant reduction in interpolation in interpolation errors with FGS and eigenvector-based, which validates the objective of these two approaches as described in section \ref{subsec:adaptive_sampling}. However, as outlined in section \ref{subsec:flops}, FGS is more efficient than eigenvector-based sampling, especially in extensive problems with large degrees of freedom, leading to more computational efficiency gain (as demonstrated in Table \ref{tab:sod_shock_eff}) and therefore is more practical and amendable for applications of large-scale problems.

In addition, the computational efficiency gains, $\lambda$, are reported in Table \ref{tab:sod_shock_eff} as the ratio of the FOM computational time to the ROM computational time, similar to Eq. \ref{eq:ROM_efficiency_gain}, but instead of evaluating the FLOPs, the FOM and ROM computational time is obtained by directly timing the corresponding online calculations.

\begin{figure}[ht]
  \centering
    \includegraphics[width=0.7\linewidth]{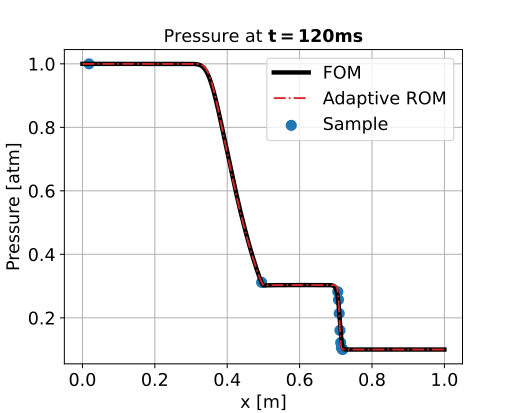} 
  \caption{Comparisons of representative FOM and adaptive ROM solutions at 120 $ms$ for the Sod shock tube}
  \label{fig:sod_shock_results} 
\end{figure}

Furthermore, we investigate the effectiveness of the feature-guided adaptive ROM in representing the physics in the Sod shock tube by comparing the instantaneous pressure solutions between FOM and adaptive ROM at $t = 120\, ms$, as depicted in Fig. \ref{fig:sod_shock_results}, which shows excellent agreement between FOM and feautre-guided adaptive ROM in accurately capturing both the advection of the shock front and the expansion of the left side of the domain. In addition, the locations of the sampling points from FGS are included to access the capabilities of the proposed sampling approach in capturing the key physics - i.e., shocks. The proposed FGS method strategically places the sample points along the trajectory of the shock front based on the features given (i.e., magnitudes of the pressure gradients), which ensures that the most crucial features (i.e., shocks) can be accurately modeled in adaptive ROM. In addition to a single shock problem, we also demonstrate the effectiveness of the feature-guided adaptive ROM in predicting a 1D problem with two colliding shocks, which is discussed in \ref{app2}.

\begin{table}[ht]
    \centering
    \caption{Comparisons of computational efficiency gain in Sod shock tube case}
    \footnotesize
    \begin{tabular}{llcccl}
        & & \multicolumn{4}{c}{\textbf{Adaptive ROM}} \\
        \cmidrule(lr){3-6}& \textbf{FOM}&\textbf{FGS}& \textbf{Random}& \textbf{Eigenvector} &\textbf{GNAT} \\
        \midrule
        \textbf{wall-clock time}&  112.6 sec&24.1 sec& 20.6 sec& 31.4 sec&26.6 sec\\
        \midrule
        \textbf{Efficiency($\lambda$)}&  -& 4.7 & 5.5 & 3.6
& 4.2 \\
    \end{tabular}
    \label{tab:sod_shock_eff}
\end{table}

\subsection{Freely propagating laminar flame}
\label{sec:flame}

Second, we demonstrate the feature-guided adaptive ROM for chemically reactive flows using a 1D, freely propagating, premixed hydorgen-oxygen laminar flame at atmospheric conditions, similar to the 1D problem created by Huang et al. \cite{huang_model_2022,huang_predictive_2023}. The 1D problem is simulated using the governing equation shown in \ref{app1}, with a detailed in $H_2-O_2$ chemical kinetic model \cite{oran_weak_1982} which involves 10 species and 24 reactions. The computational domain has a length of $10\, mm$, discretized with 1000 uniform finite volume cells, ensuring adequate resolution of the flame thickness ($\sim 1.5 \, mm$). Stoichiometric $H_2-O_2$ reactant mixture is supplied at $70.9 \, m/s$ and $315 \, K$ from the left boundary \emph{while} a constant pressure at $115 \, kPa$ is specified on the right boundary. The unsteady solution is initialized from a stationary flame profile and advanced from 0 to $50\, \mu s$ using a constant physical time step of $\Delta t = 1\, ns$ with snapshots collected at each time step. Similar to the Sod shock tube, the adaptive ROMs are constructed with different sampling methods. For consistency, all the adaptive ROMs are initialized with $0.005\, \mu s$ snapshots  ($w_{init} = 5$) from t = 0 and developed with 5 modes ($n_p = 5$ retrieving 100\% of the energy) and 12 sampling points ($n_s = 12$ covering 1.2\% of the entire computational domain), which are updated every 10 time steps ($z_s = 10$) with the first 5 sampling points selected using QDEIM while the rest 7 determined by the different oversampling strategies. Specifically, we select the magnitude of temperature as the feature for flames in FGS.

\begin{figure}[ht] 
\centering 
\includegraphics[width=0.7\linewidth]
{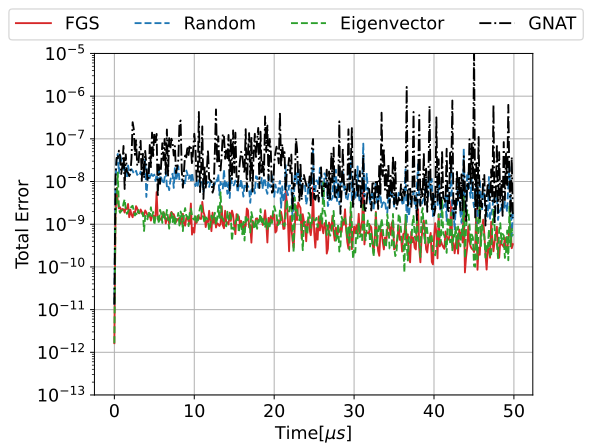} 
\caption{Comparisons of the total error measurement between the adaptive ROMs
adopting different sampling methods for the freely propagating laminar flame}
\label{fig:flame_error}
\end{figure}

\begin{table}[ht]
    \centering
    \caption{Comparisons of computational efficiency gain in freely propagating flame case}
    \footnotesize
    \begin{tabular}{llcccl}
        & & \multicolumn{4}{c}{\textbf{Adaptive ROM}} \\
        \cmidrule(lr){3-6}& \textbf{FOM}&\textbf{FGS}& \textbf{Random}& \textbf{Eigenvector} &\textbf{GNAT} \\
        \midrule
        \textbf{wall-clock time}&  182.8 min&45 min& 43.5 min& 52.8 min&44.2 min\\
        \midrule
        \textbf{Efficiency($\lambda$)}&  -&4.1& 4.2& 3.5&4.1\\
    \end{tabular}
    \label{tab:flame_eff}
\end{table}

Following the same analysis procedures in the previous section, we compute and compare the total errors (Eq. \ref{eq:total_err}) between the adaptive ROMs using the four different sampling strategies, as illustrated in Fig. \ref{fig:flame_error}, which demonstrates more than one order of magnitude accuracy improvement using FGS and eigenvector-based over random and GNAT, similar to the performance gain in the Sod shock tube in Fig. \ref{fig:sod_shock_error}. The computational efficiency gains, $\lambda$, are reported in Table \ref{tab:flame_eff}.

\begin{figure}[ht]
  \centering
    \includegraphics[width=\linewidth]{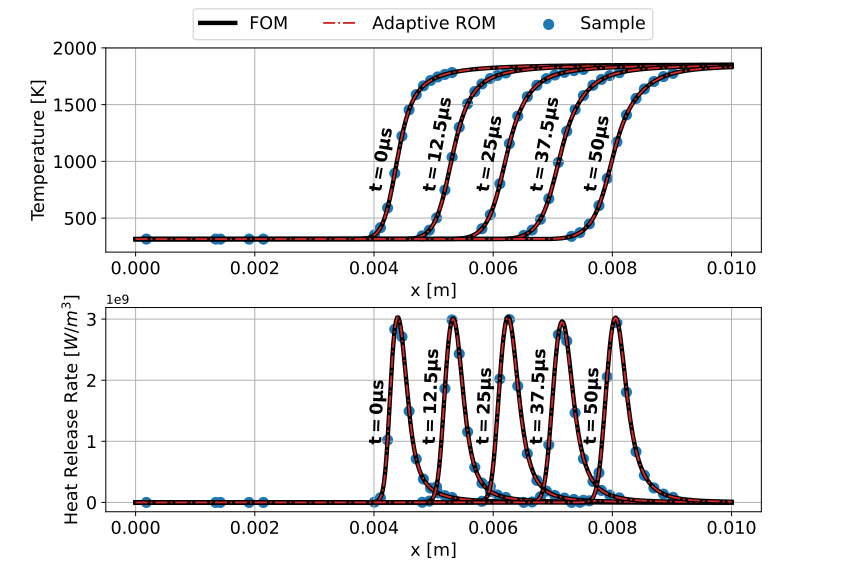}  
  \caption{Comparisons of representative 
 (a) temperature (b) heat release solutions between FOM and feature-guided adaptive ROM for the 1D freely propagating laminar flame.}
  \label{fig:flame_results} 
\end{figure}

Next, we further examine the performance of the feature-guided adaptive ROM by comparing the representative temperature and heat-release fields with FOM, as shown in Fig. \ref{fig:flame_results}, which also includes the sampling-point locations from FGS. The FOM solutions exhibit flame dynamics dominated by strong advection of sharp-gradient flame front, with temperature rise from $300\, K$ to $1850\,K$, common features indicating a slow decay of Kolmogorov N-width. Figure \ref{fig:flame_results} shows that the feature-guided adaptive ROM accurately captures the strong flame-front advection in this 1D problem with sampling points strategically populated accross the flame front by FGS. More importantly, the feature-guided adaptive ROM also accurately captures the highly nonlinear heat release rate, which is often a crucial quantity of interest for turbulent combustion simulations.

\subsection{Strongly ignited detonation tube}
\label{sec:detonation}

\begin{figure}[ht]
  \centering
    \includegraphics[width=\linewidth]{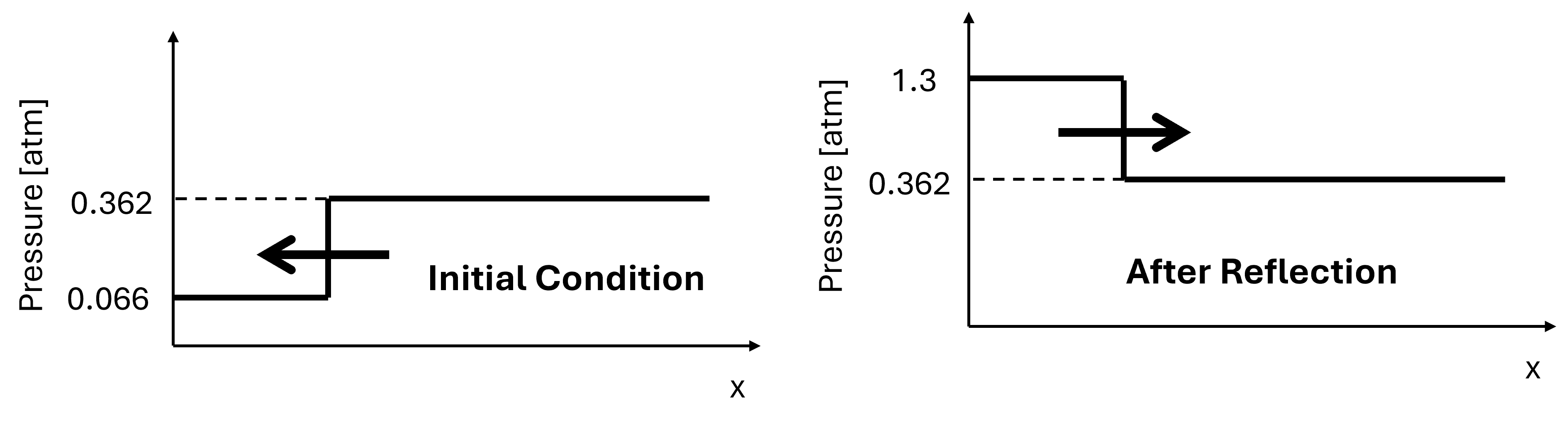} 
  \caption{Schematic of the detonation tube configuration.}
  \label{fig:digram_detonation} 
\end{figure}

\begin{figure}[ht]
  \centering
    \includegraphics[width=0.8\linewidth]{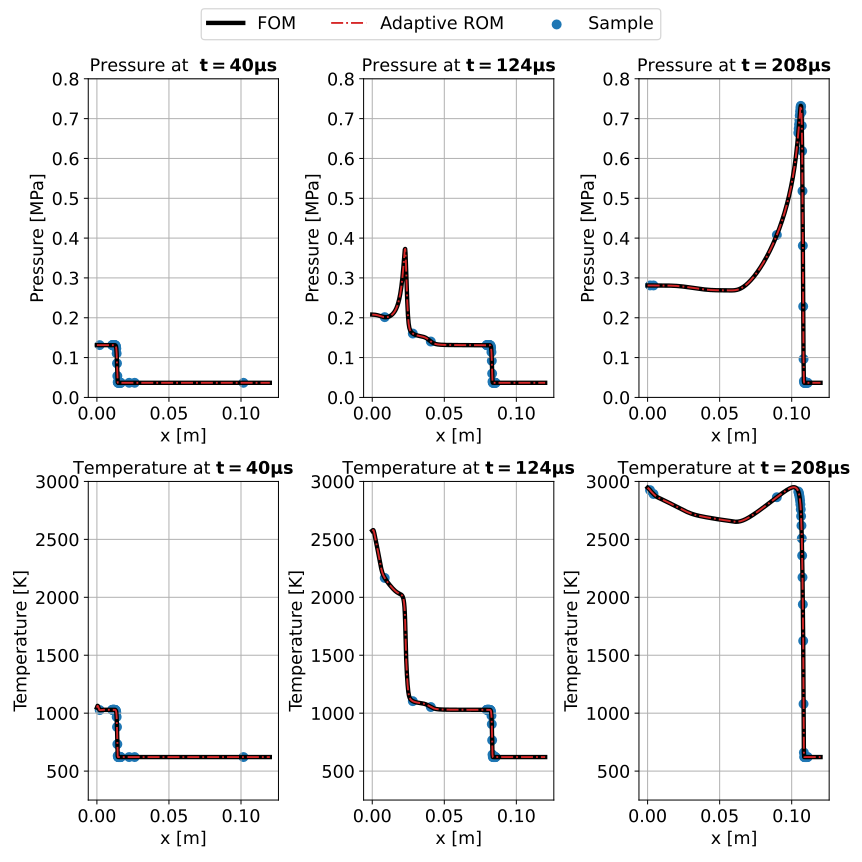}
  \caption{Comparisons of representative pressure (top) and temperature (bottom) fields after shock reflection ($t = 40\, \mu s$), during detonation-wave formation  ($t = 124\, \mu s$), and at the end of the simulation ($t = 208\, \mu s$) between FOM and feature-guided adaptive ROM for the 1D detonation tube.}
 \label{fig:detonation_results}
\end{figure}

Third, we continue the evaluations of the feature-guided adaptive ROM based on a 1D detonation tube under strong ignition of a premixed hydrogen-oxygen mixture \cite{oran_weak_1982}, which contains the challenging features of both shocks in section \ref{sec:sod_shock} and chemical reactions in section \ref{sec:flame}. The 1D detonation tube is simulated using the same $H_2-O_2$ chemical kinetics as in section \ref{sec:flame} with a computational domain spanning a length of $12 \, cm$, discretized with 2000 uniform finite volume cells. Hard wall boundary condition is applied at the left end (i.e., zero velocity $u = 0$) while the right boundary condition ensures a constant inflow of material moving at the incident shock velocity. The unsteady solution is simulated from t = 0 to  $204 \, \mu s$ using a constant physical time step size of $\Delta t = 2 \, ns$, which is determined by the extremely stiff chemical time scales under high-pressure conditions. Following the setup and observations in experiment \cite{oran_weak_1982} (summarized in Table \ref{tab:experiment_parameters}), the 1D detonation tube problem is initialized with an incident shock propagating into a quiescent $H_2-O_2-Ar$ mixture (with molar fractions ratio of 2:1:7) from the right as illustrated in Fig. \ref{fig:digram_detonation} (left). As the incident shock propagates through the mixture, the temperature and pressure increase, but not to the conditions necessary for chemical reactions to occur. Once the shock impacts the left wall and gets reflected, the temperature and pressure of the fluid increase again to the \emph{reflected} level in Table \ref{tab:experiment_parameters} (necessary to initiate chemical reactions of the specified $H_2-O_2-Ar$ mixture), as illustrated in Fig. \ref{fig:digram_detonation} (right). After an induction time, chemical reactions start and develop into a detonation wave, which moves at high speeds and quickly overtakes the shock, as shown in the FOM solutions in Fig. \ref{fig:detonation_results}.

\begin{table}[ht]
    \centering
    \caption{Parameters for the Strong Ignition}
    \footnotesize 
    \begin{tabular}{lccc}
        \toprule
         & \textbf{Undisturbed} & \textbf{Incident} & \textbf{Reflected} \\
        \midrule
        Temperature & 298 K & 621 K & 1036 K \\
        Pressure & 0.066 atm & 0.362 atm & 1.3 atm \\
        Fluid velocity & - & 4.76 $\times$ 10$^4$ cm/s & - \\
        Shock velocity & - & 7.54 $\times$ 10$^4$ cm/s & 4.5 $\times$ 10$^4$ cm/s \\
        Mach number & - & 2.165 & - \\
        \bottomrule
    \end{tabular}
    \label{tab:experiment_parameters}
\end{table}

The feature-guided adaptive ROM is initialized with $0.02\ \mu s$ of snapshots $w_{init} = 10$ from $t = 0$ and developed with 10 modes ($n_p = 10$) and 40 sample points ($n_s = 40$, corresponding to 2\% of the entire computational domain), which are updated every 8 time steps ($z_s=8$) with the first 10 sampling points selected using QDEIM while the rest 30 determined by FGS. For the detonation tube, we select the pressure gradients as the feature for shock and detonation waves in FGS. The feature-guided adaptive ROM produces gains
in computational efficiency of $\lambda \approx 6$ and the results are compared to the FOM in terms of three representative instantaneous snapshots of pressure and temperature in Fig. \ref{fig:detonation_results}, which includes one time instance after the reflection of the incident shock ($t = 40\, \mu s$), one after the detonation-wave formation ($t = 124\, \mu s$), and another one at the end simulation with a fully developed detonation wave ($t = 208\, \mu s$). 

\begin{figure}[ht]
  \centering
    \includegraphics[width=0.8\linewidth]{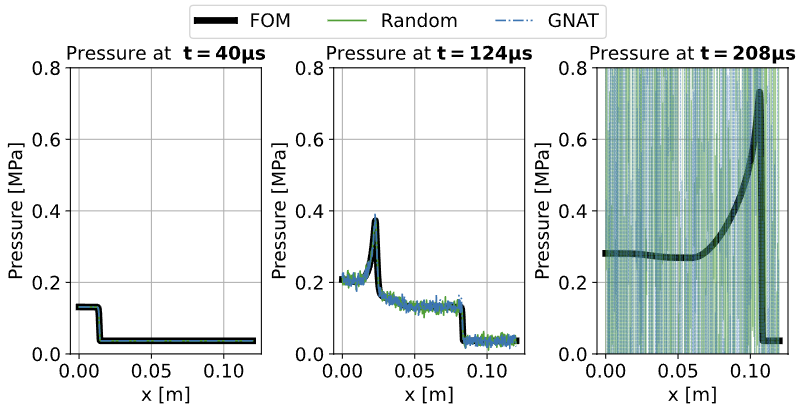}
  \caption{Comparisons of representative pressure fields after shock reflection (t = $40\ \mu s$), during detonation-wave formation  (t = $124\  \mu s$), and at the end of the simulation (t = $208 \ \mu s$) between FOM and adaptive ROMs using random and GNAT for the 1D detonation tube.}
 \label{fig:detonation_blew_up}
\end{figure}

It can be readily seen that the feature-guided adaptive ROM accurately predicts the \emph{transient} pressure and temperature dynamics present in the FOM with excellent agreement in the intensities and locations of the reflected shock and the detonation wave. We remark that the feature-guided adaptive ROM solutions match the FOM with the time-averaged total error, defined in Eq. \ref{eq:total_err}, remaining less than $10^{-10}$. Specifically, we would like to highlight that the feature-guided adaptive ROM not only successfully captures the strong advection dynamics of the detonation wave, but also precisely predicts the complex physical processes in its formation. It should be recognized that the detonation wave represents a profoundly unsteady physical phenomenon, characterized by multiple challenging features for classical ROM development such as strong convection of discontinuity prompted by the shock, advection of sharp gradients prompted by the flame, and the highly non-linear dynamics from the chemical reactions. On the other hand, directly using random and GNAT in adaptive ROM leads to unphysical solutions as depicted in Fig. \ref{fig:detonation_blew_up}, which arises when the detonation wave starts to form near $t = 40\, \mu s$. In addition, though using eigenvector-based in adaptive ROM is likely to produce accurate detonation wave predictions, it requires higher computational cost that is not scalable for large-scale problems and therefore is not considered in the current test problem. Moreover, we provide additional investigations in \ref{app3} to illustrate the performance improvements through the proposed feature-guided adaptive ROM compared to simple approaches to achieve comparable efficiency gain either through using larger physical time steps or coarser meshes.

\section{Conclusion}
\label{sec:conclusion}

Projection-based reduced-order models (ROMs) provide an efficient way to speed up high-fidelity simulations, especially in resource-intensive many-query applications like optimization and uncertainty quantification. Despite their advantages, applying ROMs to convection-dominated dynamics such as hypersonic flows and detonations poses difficulties due to the slow decay of Kolmogorov N-width, which diminishes the performance of traditional ROM methods. The current work presents a feature-guided adaptive ROM framework aimed at improving prediction accuracy in complex convection-dominated problems. It advances recent developments in adaptive model-order reduction in two ways: (1) a novel feature-guided sampling (FGS) method that selects sparse samples based strategically on key convective features of the problem, optimizing both the accuracy and efficiency of the adaptive ROMs (2) an optimized basis adaptation using multi-step snapshots collected in situ and future full-state estimation technique that incorporates non-local data to accurately predict local and global physics. This feature-focused adaptation elevates ROM precision while maintaining computational feasibility, allowing for trustworthy predictions beyond the training period. 

Comprehensive evaluations of the feature-guided adaptive ROM are performed based on a suite of 1D convection-centric problems, including (1) the Sod shock tube, (2) a freely propagating laminar flame, and (3) a multi-species detonation tube. Specifically, the FGS approach is evaluated against three established sampling methods: random, eigenvector-based, and Gauss–Newton with approximated tensors (GNAT) sampling methods. A multi-level error measurement is implemented to evaluate the accuracy of the resulting adaptive ROMs using the four different sampling methods. For the Sod shock tube, with the same level of projection errors, the FGS method is capable of producing comparable interpolation errors as the eigenvector-based method, much lower than the random and GNAT sampling methods, and therefore leading to much more accurate adaptive ROM predictions. For the freely propagating laminar flame, the feature-guided adaptive ROM successfully predicts the transient convection of the flame and provides accurate representation of the highly nonlinear heat releast rates. Furthermore, the potential of using the feature-guided adaptive ROM to predict the generation of new physical phenomena is demonstrated using the 1D detonation tube. It is shown that while the adaptive ROMs using either random or GNAT sampling methods fail to sustain its numerical stability, the feature-guided adaptive ROM successfully produces stable and robust results and more importantly, it accurately captures the formation of the detonation from a single shock wave, which is a well-recognized challenging physics for any ROM development.  Though producing similar performance in the resulting adaptive ROMs, the FGS method is demonstrated to be computationally more efficient than the eigenvector-based sampling method and is more suitable for applications of large-scale problems.

As high-performance computing becomes crucial for large-scale simulations, the feature-guided adaptive ROM framework presents a scalable and effective option for minimizing computational expenses while maintaining accuracy. Apart from optimization and uncertainty quantification, this method could also accelerate high-fidelity simulations in real-time. Future research will focus on further refining sampling strategies and frequencies, scaling for significant engineering challenges, and dynamically adjusting hyper-parameters to enhance efficiency and optimize predictive capabilities in actual applications.

\section{Acknowledgment}

The authors acknowledge the supports from the Air Force Office of Scientific Research (AFOSR) through the Center of Excellence Grant FA9550-17-1-0195 (Technical Monitors: Fariba Fahroo, Justin Koo, and Ramakanth Munipalli) and the AFOSR under the grant FA9550-
23-1-0211 (Program managers: Drs. Chiping Li and Fariba Fahroo). All the numerical studies
were carried out under the support from the Center for Research Computing at the University of
Kansas.


\appendix

\section{Governing Equations for Full Order Model}
\label{app1}

The full order model computations are carried out with an in-house one-dimensional CFD code developed in Python. This code solves the conservation equations for mass, momentum, energy and species mass fractions in a coupled fashion,

\[
\frac{\partial \mathbf{q}}{\partial t} + \frac{\mathbf{\partial f}}{\partial x}  - \frac{\mathbf{\partial f_{v}}}{\partial x} = \mathbf{H},
\]
where \(\mathbf{q}\) is the vector of conserved variables defined as,
\[
\mathbf{q} = \begin{pmatrix}
\rho \\
\rho u \\
\rho h^0 - p \\
\rho Y_l
\end{pmatrix}^T,
\]
with  $\rho$  representing density, $u$ representing velocity in x-direction, \( Y_l \) representing the \( l^{th} \) species mass fraction and the total enthalpy \( h^0 \) is defined as,
\[
h^0 = h + \frac{1}{2} u^2 = \sum_{l} h_l Y_l + \frac{1}{2} u^2.
\]
The fluxes have been separated into inviscid, \(\mathbf{F}\), and viscous terms, \(\mathbf{F}_v\), The fluxes are,
\[
\mathbf{f} = \begin{pmatrix}
\rho u \\
\rho u^2 + p \\
\rho u h^0 \\
\rho u Y_l
\end{pmatrix}, \quad
\mathbf{f}_{v} = \begin{pmatrix}
0 \\
\tau_{xx} \\
u \tau_{xx} - q_x \\
\rho D_l \frac{\partial Y_l}{\partial x}
\end{pmatrix},
\]
where \(D_l\) is defined to be the diffusion of the \(l^{th}\) species into the mixture. In practice, this is an approximation used to model the multicomponent diffusion as the binary diffusion of each species into a mixture.

The heat flux in the x-direction, \(q_x\), is defined as,
\[
q_x = -K \frac{\partial T}{\partial x} + \sum_{l=1}^N \rho D_l \frac{\partial Y_l}{\partial x} h_l + q_{\text{source}},
\]
where \( K \) is the thermal conductivity, \( T \) is the temperature, \( Y_l \) are the species mass fractions, \( h_l \) are the species specific enthalpies, and \( q_{\text{source}} \) is the heat source term.
The three terms in the heat flux represent the heat transfer due to the conduction, species diffusion and heat generation from a volumetric source (e.g. heat radiation or external heat source) respectively.

The shear stress, \(\tau\), is also found in the viscous flux and defined in terms of the molecular viscosity and velocity field,

\[
\tau_{xx} = \frac{4}{3} \mu \frac{\partial u}{\partial x},
\]
where \(\mu\) is the molecular viscosity, and $u$ is the components of the velocity field.

The source term, \(\mathbf{H}\), includes a single entry for each of the species equations signifying the production or destruction of the \(l^{th}\) species, \(\dot{\omega}_l\), which is determined by the chemical kinetics,
\[
\mathbf{H} = \begin{pmatrix} 0 & 0 & 0 & \dot{\omega}_l \end{pmatrix}^T.
\]

\section{Demonstration of Feature-Guided Adaptive ROM on Colliding Shocks}
\label{app2}

\begin{figure}[ht]
  \centering
    \includegraphics[width=0.9\linewidth]{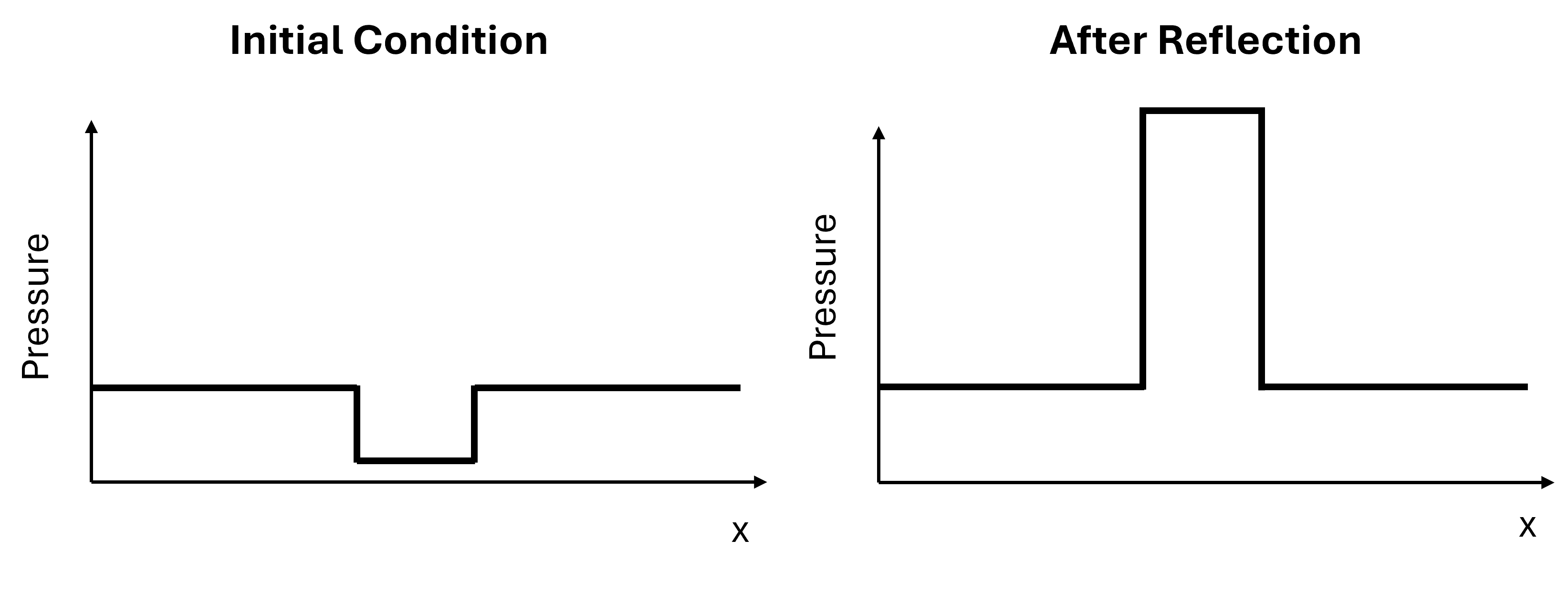} 
  \caption{Colliding Shocks Problem Setup}
  \label{fig:colliding_diagram} 
\end{figure}

In this section, we extend our evaluations further by including an extra test case featuring shock interactions using a problem with colliding shock waves. The numerical setup of the problem is shown in Fig. \ref{fig:colliding_diagram}. The computational domain is initiated with two shock waves (with pressure ratio of 6:1) traveling towards the center and colliding into each other, which leads to a substantial amplification of the shock-wave intensities with a resulting pressure ratio of almost 8:1. The computational domain has a length of 0.1 m, discretized with 1000 uniform cells. The unsteady solution is computed for $24.5\ ms$ using a constant physical time step size of $\Delta t = 7\ \mu s$.

\begin{figure}[ht]
  \centering
    \includegraphics[width=\linewidth]{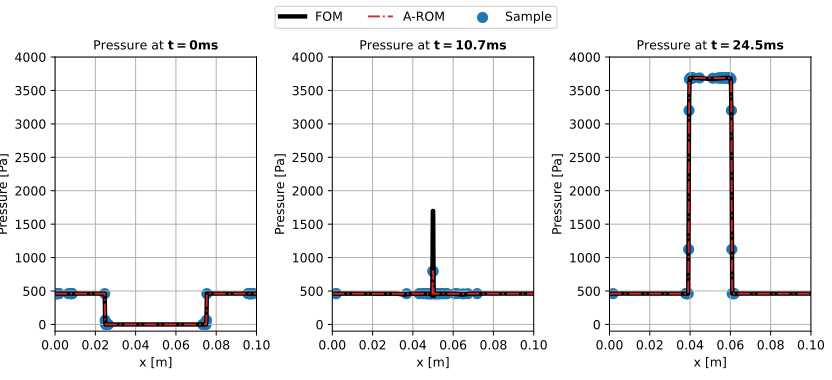} 
  \caption{Pressure Plots in Colliding Shocks Problem}
  \label{fig:colliding_results} 
\end{figure}

We develop an adaptive reduced model with an initial window size of $w_{\text{init}} = 10$ and a state dimension $n_p=9$, deploying 20 sample points within the domain accounting for 1\% of the mesh, which are refreshed every $z=10$ time steps. Figure \ref{fig:colliding_results} illustrates the pressure and sample fields at the initial condition $t = 0 \ ms$, collision moment $t = 10.7\ ms$ and at $t = 24.5\ ms$. The findings demonstrate that the FGS strategy provides an accurate approximation of the full-model pressure fields, notably offering precise localization of shock waves.

\section{Importance of adaptive ROM in Achieving Accurate Solutions Despite Accessibility of Larger Time Steps}
\label{app3}

A valid concern in the proposed algorithm in this study would be that when we are updating full state every $z_s$ physical time step in adaptive ROM, why it's necessary to develop a reduced-order model if unsteady FOM solution with a larger time step is accessible. However, the authors’ experiments demonstrate that even when comparing FOM solutions with larger time steps to those generated with the adaptive ROM, the adaptive ROM consistently delivers much more accurate results, particularly when explicit time integration methods are used. Generally, larger time steps can increase the amount of numerical diffusion because the scheme becomes less accurate in tracking sharp gradients in the solution. For example, in the case of advection or convection-dominated flows, larger time steps can cause sharp interfaces (e.g., shock waves, flames) to become smoothed out artificially due to numerical diffusion which makes larger time steps tend to accumulate larger errors. 

\begin{figure}[ht]
    \centering
    \includegraphics[width=0.5\linewidth]{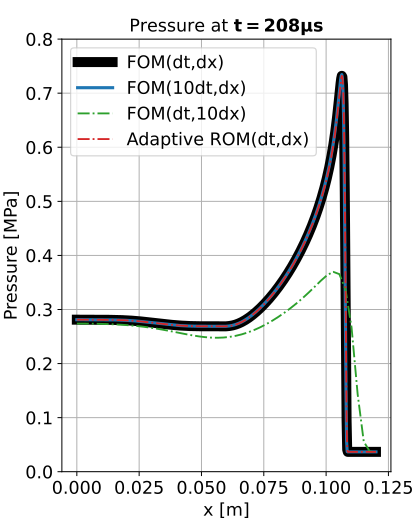}
    \caption{Adaptive ROM error analysis of strong ignition of detonation wave with larger physical time steps}
    \label{fig:detonation_larger_dt_results}
\end{figure}

To illustrate this, Fig. \ref{fig:detonation_larger_dt_results} compares results for detonation wave test case under four different scenario: (1) FOM with time step $dt = 2 \ ns$ and 2000 cells. (2) FOM with 10 times larger time step $dt = 20 \ ns$ and 2000 cells. (3) FOM with time step $dt = 2\ ns$ and 10 times coarser mesh (i.e 200 cells). (4) adaptive ROM with time step $dt = 2 \ ns$ and 2000 cells.

\begin{figure}[ht]
    \centering
    \includegraphics[width=0.8\linewidth]{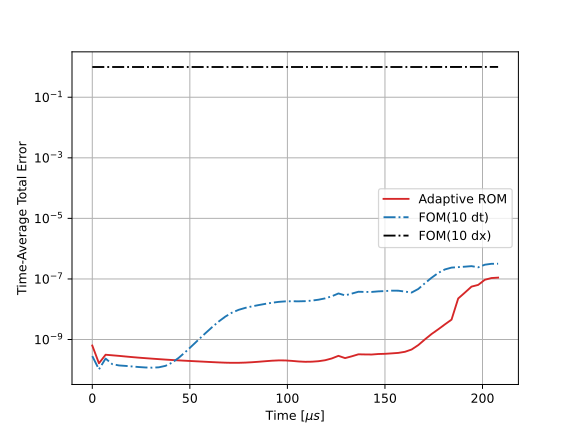}
    \caption{Adaptive ROM Error Analysis of Strong Ignition of Detonation Wave with Larger Physical Time Steps}
    \label{fig:detonation_larger_dt_error}
\end{figure}

It can be observed that, in all cases with fine meshes, the results are comparable. However, in the scenario with a coarser mesh, the detonation wave does not form, and even a sharp gradient cannot be captured. To further compare the accuracy of cases with fine meshes, the error analysis presented in Fig. \ref{fig:detonation_larger_dt_error} where the total error presented in Eq. \ref{eq:total_err} is calculated by comparing adaptive ROM with smaller time step and FOM with larger time step to FOM with smaller time step. This error analysis shows that the FOM solution with a time step 10 times larger ($dt = 20\ ns$) has a higher error compared to the adaptive ROM with a smaller time step ($dt = 2\ ns$). This demonstrates that, despite the accessibility of FOM solution with a larger time step, the adaptive ROM still provides a more reliable and precise result.

\section{Numerical Complexities of Feature-Guided and Eigenvector-Based Sampling Methods }
\label{app4}

\begin{figure}[H]
    \centering
    \includegraphics[width=0.75\linewidth]{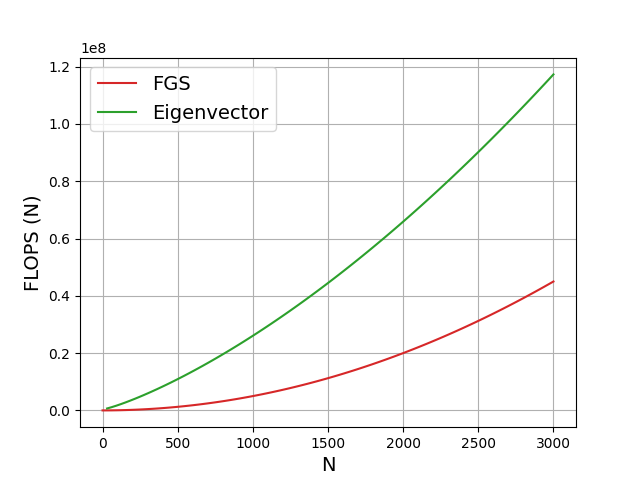}
    \caption{Comparison of FGS with eigenvector-based sampling method's numerical complexities growth rate with degrees of freedom}
    \label{fig:fgs_vs_eigen_flops}
\end{figure}

Here, we provide details on the FLOP analysis for feature-guided and eigenvector-based sampling methods, as summarized in Tables \ref{tab:FGS_FLOPS} and \ref{tab:Eigen_FLOPS} respectively. In addition, the corresponding FLOPs for the two sampling methods are compared in Fig. \ref{fig:fgs_vs_eigen_flops} against the degrees of freedom ($N$) in the target problem, which shows that as the problem size increases, the computational costs associated with the eigenvector-based method escalate more significantly compared to those of the feature-guided sampling method.

\begin{table}[ht]
    \centering
    \caption{Approximated FLOPs for feature-guided sampling}
    \footnotesize 
    \begin{tabular}{lccc}
        \toprule
         \textbf{Operations} & \textbf{Approximated FLOPs} \\
        \midrule
        QR decomposition for QDEIM & $n_pN^2$ \\
        Compute gradient of feature of interest & $2N_{elem}$ \\
        Sort the gradients & $N_{elem}log(N_{elem})$ \\
        \midrule
        Total & $n_pN^2 + 2N_{elem} + N_{elem}log(N_{elem})$ \\
        \bottomrule
    \end{tabular}
    \label{tab:FGS_FLOPS}
\end{table}

\begin{table}[ht]
    \centering
    \caption{Approximated FLOPs for eigenvector-based sampling}
    \footnotesize 
    \begin{tabular}{lccc}
        \toprule
         \textbf{Operations} & \textbf{Approximated FLOPs} \\
        \midrule
        QR decomposition for QDEIM & $n_pN^2$ \\
        Compute SVDs and sort the eigenvalues & $(n_p^3+Nlog(N))(N_{var}n_s-n_p)$ \\
        for $N_{var}n_s-n_p$ times &  \\
        \midrule
        Total & $n_pN^2 + (n_p^3+Nlog(N))(N_{var}n_s-n_p)$ \\
        \bottomrule
    \end{tabular}
    \label{tab:Eigen_FLOPS}
\end{table}

\newpage
\nocite{*}
\printbibliography
\end{document}